\documentclass[12pt,preprint,letterpaper,nofootinbib,superscriptaddress]{revtex4}
\pdfoutput=1
\usepackage[utf8]{inputenc}
\usepackage{epsf, array, xcolor}
\usepackage{graphicx}
\usepackage{subfigure}
\usepackage{bbold}
\usepackage{amsmath,amssymb,mathtools}
\usepackage
{epigraph}
\usepackage[colorlinks=true,linktocpage=true,linkcolor=blue,citecolor=blue,urlcolor=blue]{hyperref}
\usepackage{textcomp}
\usepackage{color}
\usepackage{slashed}
\usepackage{graphics,psfrag}
\usepackage{graphicx,psfrag}
\usepackage{cancel}
\def\lsim{\mbox{\raisebox{-.6ex}{~$\stackrel{<}{\sim}$~}}}

\def\dd{\text{d}}

\newcommand{\vpj }{\mbox{${\vp^\dag i\,\raisebox{2mm}{\boldmath ${}^\leftrightarrow$}\hspace{-4mm} D_\mu\,\vp}$}}
\newcommand{\vpjt}{\mbox{${\vp^\dag i\,\raisebox{2mm}{\boldmath ${}^\leftrightarrow$}\hspace{-4mm} D_\mu^{\,a}\,\vp}$}}
\newcommand{\vp}{H}

\usepackage{url}

\usepackage{dcolumn}

\begin{document}
\pagestyle{plain}

\title{The Impact of Dimension-8 SMEFT Contributions: A Case Study}

\author{Sally Dawson}
\affiliation{Department of Physics, Brookhaven National Laboratory, Upton, N.Y., 11973,  U.S.A.}
\author{Samuel Homiller}
\affiliation{Physics Department, Harvard University, Cambridge, MA, 02138, U.S.A. }
\author{Matthew Sullivan}
\affiliation{Department of Physics, Brookhaven National Laboratory, Upton, N.Y., 11973,  U.S.A.}

\date{\today}

\begin{abstract}
The use of the SMEFT Lagrangian to quantify possible Beyond the Standard Model (BSM) effects is standard in LHC and 
future collider studies.  One of the usual assumptions is to truncate the expansion with the dimension-$6$ operators.  The numerical impact of the
next terms in the series, the dimension-$8$ operators, is unknown in general.  We consider a specific BSM model containing a charge-${2 / 3}$ heavy vector-like quark
 and compute the operators generated at dimension-$8$.  The numerical effects of these operators are studied for the $t\bar{t}h$
process, where they contribute at tree level and we find effects at the ${\cal{O}}(0.5-2\%)$ level for allowed values of the parameters. 
\end{abstract} 

\maketitle


\section{Introduction} 

One of the goals of the HL-LHC running is a precision physics program that enables a detailed comparison
of theoretical and experimental predictions.  
Lacking the experimental discovery of any new particles, the tool of choice is the Standard Model Effective
Field Theory (SMEFT) which assumes that  the gauge symmetries and particles of the Standard Model provide an
approximate description of weak scale physics~\cite{Brivio:2017vri}.  Deviations from the Standard Model (SM) predictions are parameterized in terms of
an infinite tower of higher dimension operators,
\begin{equation}
\mathcal{L} \sim \mathcal{L}_{\textrm{SM}}+\sum_n \sum_i {C_i^{(n)} \over \Lambda^{n-4}}\, O_i^{(n)}, 
\end{equation}
where $\Lambda$ is a high energy scale where some unknown UV complete model is presumed to exist.  All of the new
physics information resides in the coefficient functions, $C_i^{(n)}$, which can be extracted from experimental data. 

The SMEFT amplitude for a tree level scattering process can be written schematically as,
\begin{equation} 
\mathcal{A} \sim \mathcal{A}_{\textrm{SM}} +\sum_i  {C_i^{(6)} \over \Lambda^2} \mathcal{A}_i^{(6)} 
+\sum_i  {C_i^{(8)} \over \Lambda^4} \mathcal{A}_i^{(8)} +...
 \, ,
\end{equation}
where $\mathcal{A}_{\textrm{SM}}$, $\mathcal{A}_i^{(6)}$ and $\mathcal{A}_i^{(8)}$ are the SM, dimension-$6$ and dimension-$8$ contributions, 
respectively\footnote{We neglect baryon and lepton number violating operators.}. Squaring the amplitude, a
physical cross section takes the form of an integral over the appropriate phase space, $\dd {\rm PS}$,
\begin{eqnarray}
\dd \sigma &\sim &\int (\dd {\rm PS})\biggl\{ \mid \mathcal{A}_{\textrm{SM}}\mid ^2 +{2\over \Lambda^2} 
\textrm{Re}\biggl(\sum C_i^{(6)} \mathcal{A}_i ^{(6)} \mathcal{A}_{\textrm{SM}}^*\biggr)
+{1\over \Lambda^4}
\textrm{Re}\biggl(\sum_{i, j}
C_i^{(6)}C_j^{(6)*} \mathcal{A}_i^{(6)} \mathcal{A}_j^{(6)~*}\biggr)
\nonumber \\
&&+{2\over \Lambda^4} \textrm{Re} \biggl(\sum_i C_i^{(8)} \mathcal{A}_i^{(8)} \mathcal{A}_{\textrm{SM}}^*\biggr)\biggr\}\, + \dots 
\end{eqnarray}
It is immediately apparent that the squares of the dimension-$6$ contributions are formally of the same power counting in ${1 /\Lambda^4}$
as the interference of the dimension-$8$ terms with the SM result unless assumptions are made about the
relative sizes of the contributions.  If the process being studied is extremely well constrained (as is the 
case for the electroweak precision observables), it may be sufficient to include only the ${1 / \Lambda^2}$ contributions, as the ${1 / \Lambda^4}$
terms are negligible in this case~\cite{Dawson:2019clf,Ethier:2021bye,Almeida:2021asy}.  Alternatively, the SMEFT could result from a strongly interacting theory
at the UV scale where the ${\mathcal{A}_{\textrm{SM}} \mathcal{A}^{(8)} / \Lambda^4}$ terms are suppressed relative to 
the ${| \mathcal{A}^{(6)} |^2 / \Lambda^4}$ contributions~\cite{Contino:2016jqw,Trott:2016}.  There are, however, scenarios where the inclusion of the
dimension-$8$ terms  may be critical  in order to obtain reliable results due to cancellations of the ${|\mathcal{A}^{(6)}|^2 / \Lambda^4}$ 
terms in specific kinematic regimes~\cite{Panico:2017fr}. There are also scenarios where new physics effects first arise at dimension-8 such as the $ZZ\gamma$ 
coupling\cite{Buchmuller:1985jz,ZZ:2020}. Furthermore, in weakly coupled theories, there is generically no reason to
expect the dimension-$8$ contributions to be suppressed. 

In practice, the SMEFT series is usually terminated at dimension-$6$ and the amplitude is computed to ${\cal{O}}({1 /\Lambda^2})$, generating
${\cal{O}}({1 / \Lambda^4})$ contributions in cross sections. This leaves an uncertainty about the numerical relevance of the higher dimension operators.  A complete basis for the dimension-$8$ operators
now exists~\cite{Hays:2018zze,Murphy:2020ab,Murphy:2020cly,Li:2020tsi,Lin:2021}, making possible phenomenological studies of the effects of these operators. 
The literature, however,  contains very few concrete examples of the effects of dimension-$8$ contributions.  
Studies of a subset of dimension-$8$ contributions to Higgs plus jet production show a modest distortion of kinematic shapes at high $p_T$~\cite{Dawson:2015gka,Dawson:2014ora,Harlander:2013oja,Grazzini:2016paz,Battaglia:2021nys}. Ref.~\cite{Hays:2018zze}  considers the dimension-8 contributions
to $Wh $ production and notes that quite large cancellations  between the contributions of different  dimension-$8$ operators are possible.  
In a similar vein, the authors of Ref.~\cite{Corbett:2021eux} compute $Z$ pole observables
to ${\cal{O}}({1 / \Lambda^4})$ and  find numerically significant effects.  These examples consider the SMEFT coefficients as arbitrary unknown
parameters.  In a given UV model, however, the coefficients are predicted, and the conclusions that can be drawn from studies of SMEFT parameters depend 
sensitively on the relationships between the different coefficients at the UV scale~\cite{Dawson:2020oco,Almeida:2021asy,Brivio:2021alv}.

 In this paper, we discuss an example
of UV physics where the coefficients of the dimension-$6$ and dimension-$8$ operators can be computed in terms of a small number 
of input parameters, allowing us to assess the relevance of terms of ${\cal{O}}({1 / \Lambda^4})$ arising from the dimension-$8$ operators. 
The example we consider contains a charge-${2 / 3}$ vector like top quark (TVLQ) that is assumed to exist at the UV scale.   
Such particles occur in little Higgs models~\cite{Arkani-Hamed:2002ikv,Perelstein:2003wd,Csaki:2002qg}
and in many composite Higgs models~\cite{Panico:2015jxa,Matsedonskyi:2015dns,Dobrescu_1998,He_2002}, and represent a highly motivated scenario.  Within the context of this model, the coefficients of the dimension-$6$ and
dimension-$8$ operators can be calculated  using the covariant derivative expansion~\cite{Gaillard:1985uh,Henning:2014wua} and matched to the SMEFT. This allows for 
a detailed numerical analysis of the
various approximations frequently used when computing observables in the SMEFT. We consider $t \bar{t} h $ associated production in the SMEFT limit  of the TVLQ and  are able to concretely determine the numerical relevance of
the dimension-$8$ contributions to this process at tree level. The SM rate for $tth$ production at the LHC is well known at NLO QCD~\cite{Beenakker:2001rj,Beenakker:2002nc,
Dawson:2002tg,Dawson:2003zu}.

In Sec.~\ref{sec:modeldef}, we review the construction of the TVLQ model and 
we pay particular attention to the decoupling properties of the TVLQ model.
The tree-level matching to the SMEFT at dimension-$8$ is given in Sec.~\ref{sec:smeft_match}.
Phenomenological results for $t {\overline t}h $ at dimension-$8$ in the SMEFT limit of the TVLQ are presented in Sec.~\ref{sec:tthpheno}, where we emphasize the importance of including the top decay products for SMEFT studies.  
We conclude with a discussion of the impact of our results in Sec.~\ref{sec:discussion}.
Appendices include a short summary of the relevant dimension-8 interactions and a brief discussion of one-loop matching in the TVLQ model.

\section{The TVLQ Model}
\label{sec:modeldef}

We consider an extension of the Standard Model with one additional vector-like, charge-${2 / 3}$ quark, denoted $\mathcal{T}_L^2$, $\mathcal{T}_R^2$, that can mix with the Standard Model-like 
top quark, $\mathcal{T}_L^1,\mathcal{T}_R^1$ and call this the TVLQ model.  This model has been extensively studied in the 
literature~\cite{Cacciapaglia:2010vn, Chen:2017hak, Cacciapaglia:2018qep, Buchkremer:2013bha, Cacciapaglia:2011fx, Matsedonskyi:2014mna, delAguila:1998tp, Aguilar-Saavedra:2002phh, Aguilar-Saavedra:2013qpa, Ellis:2014dza, Aguila_2000, Chen:2014xwa, Buchkremer_2013} and we briefly summarize the 
salient points. 
The SM-like third generation chiral fermions are,
\begin{equation}
\psi_L = \begin{pmatrix} \mathcal{T}_L^1 \\ b_L \end{pmatrix},
\quad
\mathcal{T}_R^1,
\quad
b_R\, ,
\end{equation}
with the usual Higgs Yukawa couplings:
\begin{equation}
\mathcal{L}_{\textrm{Yuk}}^{\textrm{SM}} = 
- \lambda_b \bar{\psi}_L H b_R 
- \lambda_t \bar{\psi}_L {\tilde H} \mathcal{T}_R^1+ \textrm{h.c.}\, ,
\label{eq:yukawas_sm}
\end{equation}
where ${\tilde H}_i=\epsilon_{ij}H^*_j $.
Note that we will distinguish between the SM-like Yukawa couplings, $\lambda_b, \lambda_t$ in Eq.~\eqref{eq:yukawas_sm}, their SM values, $Y_b = {\sqrt{2} m_b / v}$, $Y_t = {\sqrt{2} m_t / v}$ with $m_b$ and $m_t$ the physical quark masses, and the Yukawa couplings derived in the SMEFT construction of Sec.~\ref{sec:smeft_match}.
As usual, $v=(\sqrt{2}G_F)^{-1/2}$.

The most general fermion mass terms for the charge-${2 / 3} $ quarks are:
\begin{equation}
\mathcal{L} = \mathcal{L}_{\textrm{Yuk}}^{\textrm{SM}} 
- \lambda_{T} \bar{\psi}_L H^c \mathcal{T}_R^2 
- m_{12} \bar{\mathcal{T}}_L^2 \mathcal{T}_R^1 
- m_{\mathcal{T}} \bar{\mathcal{T}}_L^2 \mathcal{T}_R^2 
+ \textrm{h.c.}
\end{equation}
Since $\mathcal{T}_R^2$, $\mathcal{T}_R^1$ have identical quantum numbers, the $m_{12}$ term can be set to zero by a redefinition of the fields. The charge-${2 / 3}$ sector is thus described by
three parameters: $\lambda_t,~\lambda_T$ and $m_{\mathcal{T}}$. 

The physical fields, $t$ and $T$, with masses $m_t$ and $M_T$, are found by diagonalizing the mass matrix with two unitary matrices, 
\begin{eqnarray}
\begin{pmatrix}
t\\
T
\end{pmatrix}_{L,R}
=
\begin{pmatrix} 
\cos\theta_{L,R} & -\sin\theta_{L,R}\\
\sin\theta_{L,R} & \cos\theta_{L,R}
\end{pmatrix}
\begin{pmatrix}
\mathcal{T}_L^1\\
\mathcal{T}_L^2
\end{pmatrix}_{L,R}
\end{eqnarray}
and we use the shorthand $ c_{L,R} \equiv\cos\theta_{L,R}$, and $s_{L,R}\equiv\sin\theta_{L,R}$.

\begin{figure}
\centering
\includegraphics[width=0.47\textwidth]{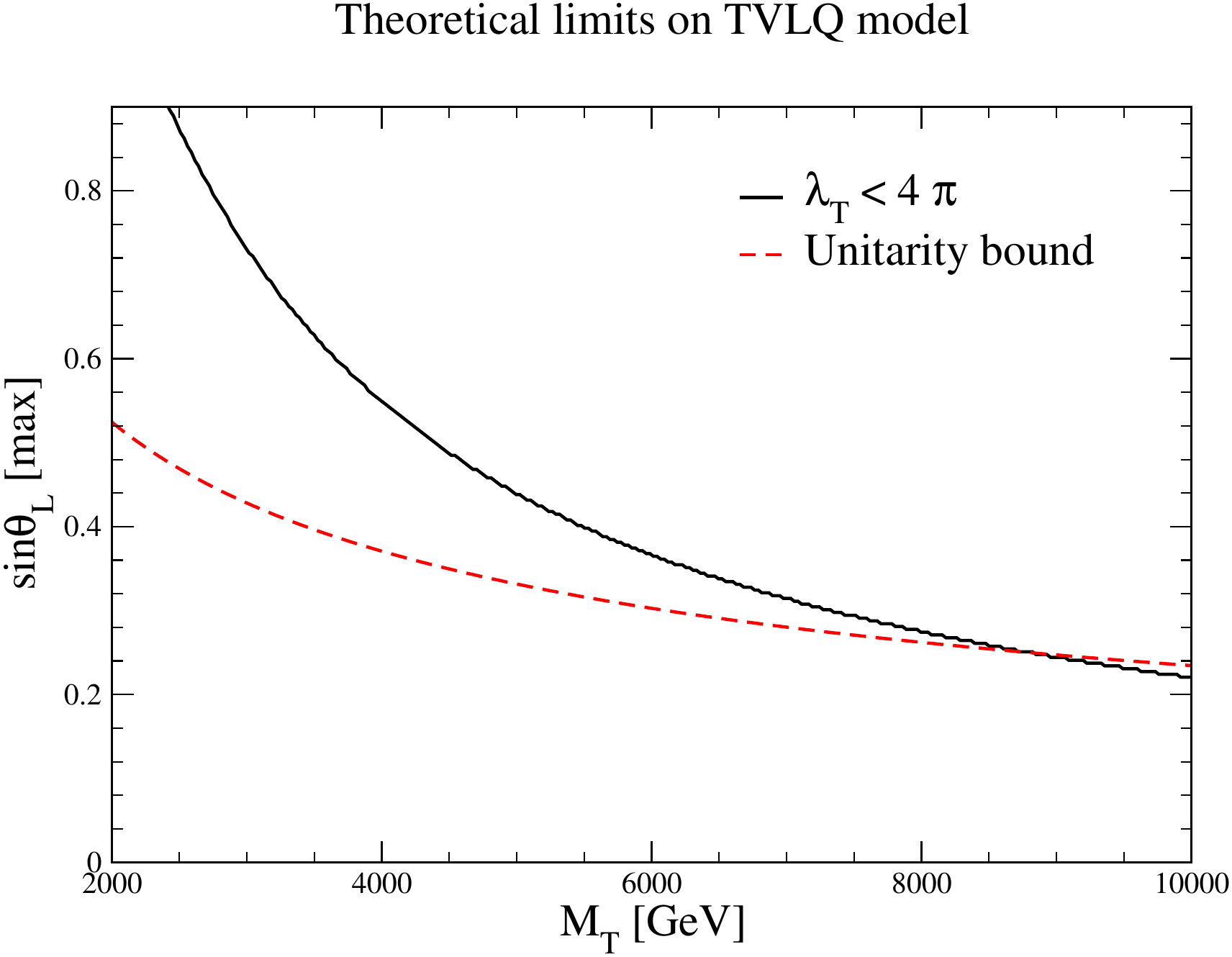}
\includegraphics[width=0.47\textwidth]{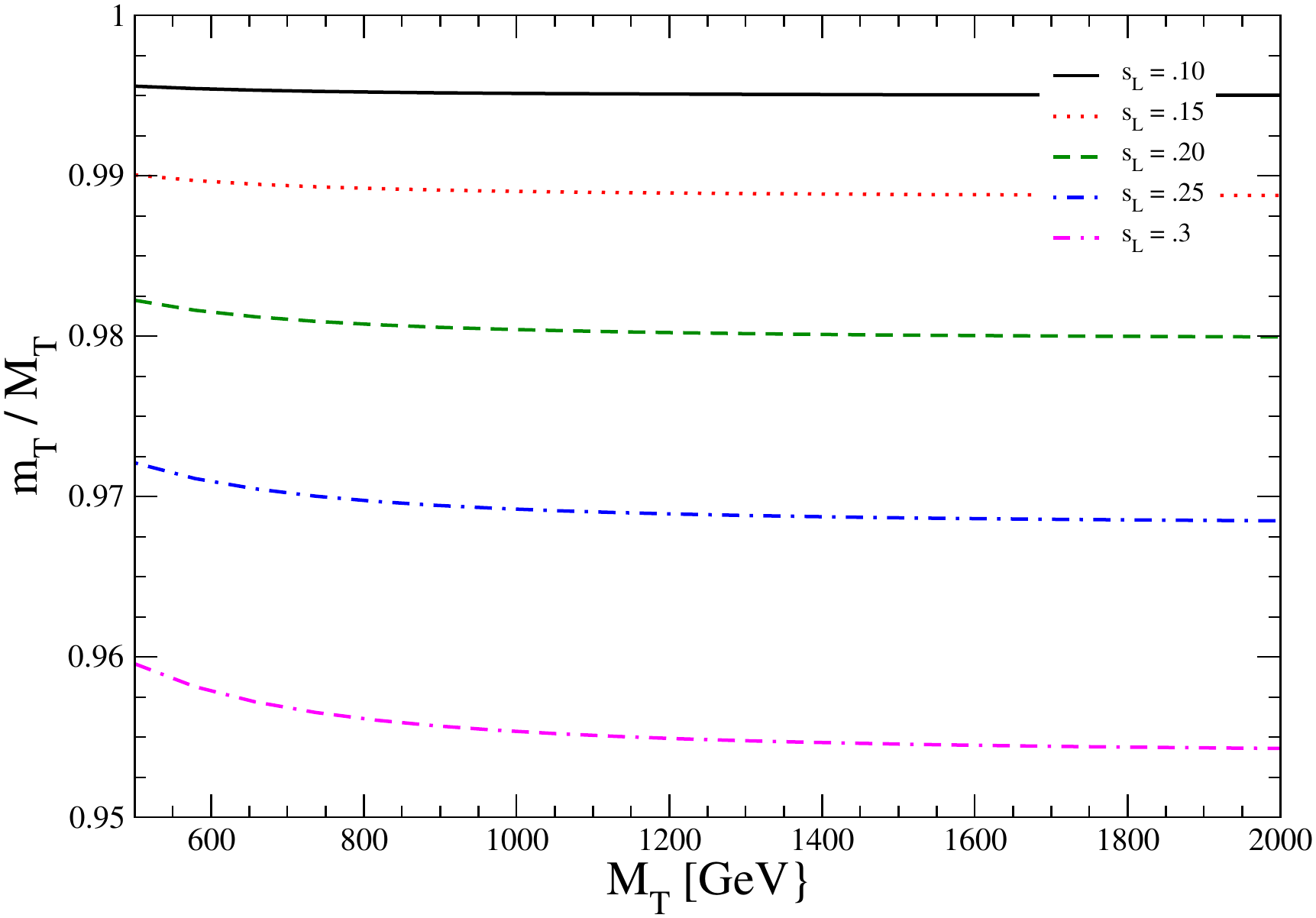}
\caption{LHS: Upper limit on $s_L$ as a function of $M_T$ from perturbativity and unitarity.
RHS: Relationship between the Lagrangian mass, $m_{\mathcal{T}}$, and the physical mass, $M_T$, for fixed $\sin\theta_L\equiv s_L$.}
\label{fig:dec2}
\end{figure}

Useful relationships between the Lagrangian and physical parameters are, 
\begin{eqnarray}
{\lambda_t v\over\sqrt{2}}&=& c_L c_R m_t+s_Ls_R M_T={s_R\over s_L}M_T\nonumber \\
{\lambda_T v\over \sqrt{2}}&=& -s_R c_L m_t +s_L c_R M_T={s_Rc_L\over m_t}(M_T^2-m_t^2)\nonumber\\
 m_{\mathcal{T}}&=& s_L s_R m_t+c_Lc_RM_T={s_R\over s_Lm_t}(s_L^2m_t^2+c_L^2M_T^2)\nonumber\\
 \tan \theta_R&=& {m_t\over M_T}\tan\theta_L\, .
 \label{eq:rels}
 \end{eqnarray}
 The following relationships follow from Eq.~\eqref{eq:rels},
 \begin{eqnarray} 
 {\lambda_T v\over\sqrt{2}m_{\mathcal{T}}}
& =&{c_Ls_L(1-x)\over \sqrt{1-s_L^2(1-x)}}\nonumber \\
 {\lambda_t v\over\sqrt{2}}&=& {m_t \over \sqrt{1-s_L^2(1-x)}}
 \nonumber \\
 {\lambda_Tv\over\sqrt{2}}&=& M_T{s_L c_L (1-x)\over \sqrt{(1-s_L^2(1-x))}}\nonumber \\
 m_{\mathcal{T}}&=& M_T\sqrt{1-s_L^2(1-x)}\, ,
 \label{eq:exact}
 \end{eqnarray}
 with $x \equiv {m_t^2 / M_T^2}$.
 From Eq.~\eqref{eq:exact}, it is clear that  for fixed $s_L$, $\lambda_T$ will become non-perturbative at large $M_T$.  In Fig.~\ref{fig:dec2} (LHS) ,
 we show the upper limit on $s_L$ from the requirement that $\lambda_T\lsim 4 \pi$, along with the unitarity
 limit from $\overline{T}T\rightarrow \overline{T}T$ of $s_L^2 \lsim 550\,\textrm{GeV}/M_T$~\cite{Chanowitz:1978mv}.
 The $M_T\rightarrow\infty$ limit therefore requires $s_L\rightarrow 0$ for a weakly interacting theory.  
 We also observe that the expansions in $1/M_T^2$ and $1/m_{\mathcal{T}}^2$
 have different counting in inverse mass dimensions for fixed $s_L$, 
\begin{eqnarray}
{1\over m_{\mathcal{T}}^2}&=& {1\over M_T^2}+{s_L^2\over M_T^2}\biggl(1-{m_t^2\over M_T^2}\biggr)\, ,
\end{eqnarray}
as is demonstrated in Fig.~\ref{fig:dec2} ~(RHS). 
The ratio $m_{\mathcal{T}}/M_T$ quickly goes to its asymptotic limit as $M_T\rightarrow \infty$ and for $s_L\sim 0.2$, the ratio approaches $\sim 0.98$, for example. 

The relations of Eq.~\eqref{eq:exact} can be inverted~\cite{Cacciapaglia:2010vn}:
\begin{eqnarray}
m_t^2&=&
{(\lambda_t^2+\lambda_T^2)v^2\over 4}
+{m_{\mathcal{T}}^2\over 2}
\biggl[1-\sqrt{\biggl(1+{(\lambda_t^2+\lambda_T^2)v^2\over 2
m_{\mathcal{T}}^2}\biggr)^2-{2\lambda_t^2 v^2\over m_{\mathcal{T}}^2}}\biggr]
\nonumber \\
M_T^2&=& {m_{\mathcal{T}}^2\over m_t^2}\biggl({\lambda_t^2 v^2\over 4}\biggr)\nonumber \\
s_L&=& {\lambda_T v\over \sqrt{2} m_{\mathcal {T}}}
{
1\over \sqrt{ 
(1-{m_t^2\over m_{\mathcal{T}}^2})^2+
{\lambda_T^2 v^2\over 2 m_{\mathcal{T}}^2}
}} \, .
\end{eqnarray}
In our phenomenological studies we will switch between Lagrangian parameters and the physical
parameters to illustrate various points.  We remind the reader that the physical masses are $m_t$ and $M_T$ with $m_t << M_T$ and that $m_{\mathcal{T}}$ is the Lagrangian parameter. 



\begin{figure}
\centering
\includegraphics[width=0.47\textwidth]{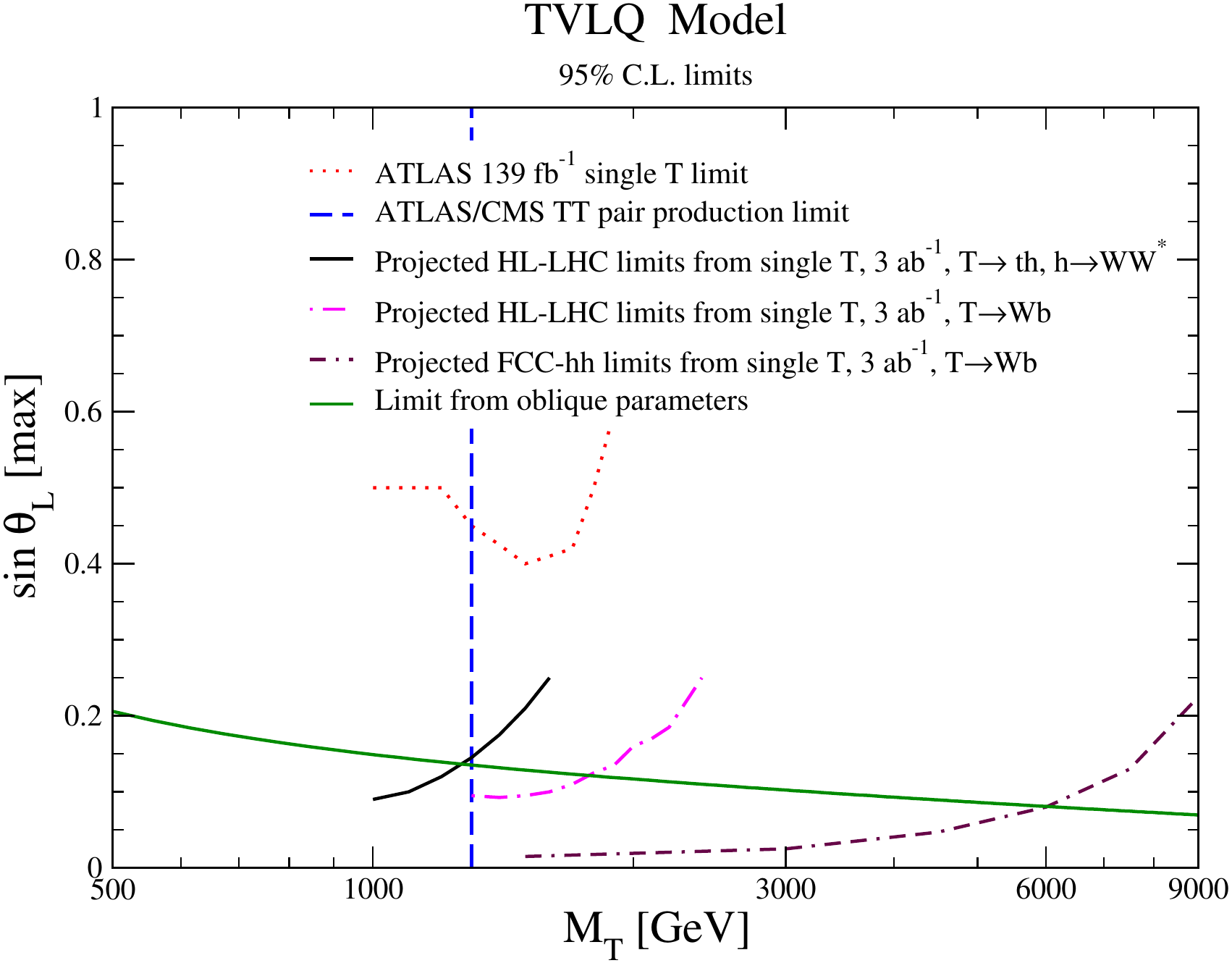}
\caption{
LHC $95\% $ exclusion limits on TVLQ masses from single $T$ production with $T\rightarrow ht, Zt$ as a function of $s_L$~\cite{ATLAS_singt} (red dotted line) and limits from $T{\bar T}$ production with $T\rightarrow th, Zt, Wb$~\cite{ATLAS:2018ziw,cms_2018} (blue dashed), projected limits from single $T$ production at HL-LHC with $T\rightarrow th,h\rightarrow W W^*$~\cite{Liu_2019}, (black solid), projected limits at HL-LHC from single $T$ production with $T\rightarrow Wb$~\cite{Yang:2021btv}(magenta dot-dashed), and projected limits at FCC-hh ($\sqrt{S}=100\,\textrm{TeV}$) from single $T$ production with $T\rightarrow Wb$~\cite{Yang:2021btv}(purple long dot-dashed). The region above the curves is excluded.}
 \label{fig:stu}
\end{figure}

The oblique parameters place stringent limits on the parameters of the TVLQ.  In Fig.~\ref{fig:stu}, we update the results of Ref.~\cite{Chen:2017hak}, include the global fit results of Ref.~\cite{Dawson:2020oco} and compare with the direct search limits from $T\bar{T}$ pair production~\cite{ATLAS:2018ziw,cms_2018} (which are independent of $s_L$).  We also show a comparison of current searches with projections for HL-LHC and FCC-hh and note that the HL-LHC will be sensitive to $M_T\sim 1.7~\textrm{TeV}$, while 
the FCC-hh can probe up to $\sim 6~\textrm{TeV}$ \cite{Liu_2019, Yang:2021btv}\footnote{We note that $g^*=2s_L$.}.

\section{Matching to SMEFT at dimension-$8$}
\label{sec:smeft_match}

In this section, we consider the $M_T\rightarrow \infty$ limit of the TVLQ model and perform the tree-level matching to the SMEFT, extending
the dimension-$6$ results~\cite{Aguila_2000,Chen:2014xwa,Buras_2011}  to dimension-$8$. 
Since the full UV model depends on only three unknown parameters, it is particularly simple. 
We use the covariant derivative expansion~\cite{Gaillard:1985uh,Henning:2014wua} to integrate the heavy $T$ out  of the theory and generate the effective operators at dimension-$6$ and dimension-$8$.
The resulting Lagrangian involving the SM-like top quark, $t$, is,
\begin{equation}
\mathcal{L}_t \equiv \mathcal{L}_{\textrm{kin}} + \mathcal{L}_Y + \mathcal{L}_6 + \mathcal{L}_8\, ,
\end{equation}
where,
\begin{eqnarray}
\mathcal{L}_{\textrm{kin}}&=& i{\overline \psi}_L\slashed{D}\psi_L + i{\overline t}_R\slashed{D}t_R\nonumber \\
&=&{i\over 2} \biggl[{\overline \psi}_L (\slashed{D}\psi_L)-( {\overline \psi}_L \overleftarrow{\slashed{D}} ) \psi_L\biggr]+ i{\overline t}_R\slashed{D}t_R
\nonumber \\ 
\mathcal{L}_Y&=&-\lambda_t {\overline \psi}_L {\tilde H}t_R + \textrm{h.c.} \nonumber\\ 
\mathcal{L}_6&=&
 \frac{i}{2}{\lambda_T^2\over m_{\mathcal{T}}^2} 
{\overline{\psi}}_L{\tilde{H}}\slashed{D} ({\tilde{H}}^ \dagger\psi_L)  +  \textrm{h.c.}
\nonumber \\
\mathcal{L}_8&=&
-\frac{i}{2} {\lambda_T^2\over m_{\mathcal{T}}^4} 
{\overline{\psi}}_L{\tilde{H}}\slashed{D}^3 ({\tilde{H}}^ \dagger\psi_L)  +  \textrm{h.c.} ,
\label{eq:theory}
\end{eqnarray}
where $\slashed{D} \equiv \gamma^\mu D_\mu$. The covariant derivative $D_\mu$ is defined as $\partial_\mu -i \frac{g}{2} Y B_\mu - i g_W \tau_a W^a_\mu -i g_s T_a G^a_\mu$, with hypercharge $Y$, $SU(2)$ generators $\tau_a$, and $SU(3)$ generators $T_a$. 
The dimension-$6$ term, $L_6$,  generates a non-standard normalization for the  top quark kinetic energy term after electroweak symmetry breaking and the expansion of the Higgs field around its vev, so we make the  gauge invariant field redefinition~\cite{Criado:2018sdb,Brivio:2017bnu},
\begin{eqnarray}
\psi_{L,i}&\rightarrow & \biggl(\psi_{L,i}-{\lambda_T^2\over  2  m_{\mathcal{T}}^2}(\tilde{H}^{\phantom{\dagger}}_i \tilde{H}^\dagger_j) \psi_{L,j}\biggr)\, ,
\label{eq:fieldr}
\end{eqnarray}
where $i, j$ are $SU(2)$ indices.  This brings the top quark kinetic energy into the canonical form. 

\begin{table}[t]
\centering
\begin{tabular}{||c|c||}
\hline\hline
Dimension-$6$ & \\
\hline\hline
$O_{Ht}^{1,(6)}$ &  $(\vpj)(\bar \psi_L  \gamma^\mu \psi_L)$ \\
\hline
$O_{Ht}^{3,(6)}$ & $(\vpjt)(\bar \psi_L \tau^a \gamma^\mu \psi_L)$  \\
\hline
$O_{tH}^{(6)}$  & $(H^\dagger H) {\overline \psi}_L {\tilde{H}} t_R$ \\
\hline\hline
Dimension-$8$ & \\
\hline\hline 
$O_{quH^5}^{(8)}$ & $(H^\dagger H) ^2{\overline \psi}_L {\tilde H} t_R$ \\
\hline\hline
\end{tabular}
\caption{Operators appearing in Eq.~\eqref{eq:newl6} where $\psi_L$ is the $(t,b)$ doublet,
 $H^\dagger   \raisebox{2mm}{\boldmath ${}^\leftrightarrow$}\hspace{-4mm} D_\mu\,H\equiv H^\dagger D_\mu H-(D_\mu H^\dagger) H$ and 
$H^\dagger   \raisebox{2mm}{\boldmath ${}^\leftrightarrow$}\hspace{-4mm} D_{\mu}^ a\,H\equiv H^\dagger D_\mu\sigma^a H-(D_\mu   H^\dagger \sigma^a) H$.\label{tab:opdef}}
\end{table}

After making the  field redefinition of Eq.~\eqref{eq:fieldr},
\begin{eqnarray}
i{\overline \psi}_L\slashed{D}\psi_L+ \mathcal{L}_6 &\rightarrow & 
i{\overline \psi}_L\slashed{D}\psi_L +{i\lambda_T^2\over 2 m_{\mathcal{T}}^2}
\biggl[ ({\overline \psi}_L {\tilde H}) (\slashed{D}{\tilde H}^\dagger )\psi_L
-{\overline \psi}_L(\slashed{D}{\tilde H}) {\tilde H}^\dagger \psi_L\biggr]+\delta \mathcal{L}_{8a}\nonumber \\
 &= &
i{\overline \psi}_L\slashed{D}\psi_L +{i\lambda_T^2\over 4 m_{\mathcal{T}}^2}\biggl[
 ({\overline \psi}_L \gamma^\mu \psi_L)
 \biggl(H^\dagger   D_\mu H-(D_\mu H^\dagger) H\biggr)\nonumber \\ &&
  -({\overline \psi}_L \gamma^\mu\tau^a \psi_L)
 \biggl(H^\dagger  \tau^a D_\mu H
 -(D_\mu H^\dagger)\tau^a H\biggr)
\biggr]+\delta \mathcal{L}_{8a}\nonumber \\
&= &
i{\overline \psi}_L\slashed{D}\psi_L +{i\lambda_T^2\over 4 m_{\mathcal{T}}^2}\biggl[O_{Ht}^{1,(6)}
-O_{Ht}^{3, (6)}\biggr]+
\delta \mathcal{L}_{8a}\nonumber \\
\delta \mathcal{L}_{8a} &\equiv & {i\lambda_T^4\over 8 m_{\mathcal{T}}^4} \biggl[ 
3\slashed{D} ({\overline\psi}_L{\tilde H}){\tilde H}^\dagger  {\tilde H} {\tilde H}^\dagger \psi_L
+{\overline\psi}_L{\tilde H}{\tilde H}^\dagger (\slashed{D}{\tilde H}) {\tilde H}^\dagger \psi_L+ \textrm{h.c.} \biggr]
\nonumber \\
&=&  {\lambda_T^4\over 8 m_{\mathcal{T}}^4}
\biggl[ -3  \lambda_t O_{quH^{5}}^{(8) }
+3(H^\dagger H) {\overline  \psi_L} ( i\slashed{D} {\tilde H}){\tilde H}^\dagger \psi_L 
+{\overline{\psi}}_L{\tilde H} {\tilde H}^\dagger ( i\slashed{D} {\tilde H}){\tilde H}^\dagger \psi_L +
\textrm{h.c.} \biggr]\nonumber \\ 
\mathcal{L}_Y&\rightarrow& -\lambda_t {\overline \psi}_L {\tilde H} t_R \biggl[1-{\lambda_T^2\over 2 m_{\mathcal{T}}^2} (H^\dagger H) \biggr] + \textrm{h.c.} \nonumber \\ 
&\equiv  & \biggl[-\lambda_t {\overline \psi} _L{\tilde H} t_R  +{\lambda_t\lambda_T^2\over 2 m_{\mathcal{T}}^2}O_{tH}^{(6)} + \textrm{h.c.} \biggr] \, , 
\label{eq:newl6}
\end{eqnarray}
where in the second line of Eq.~\eqref{eq:newl6}, $SU(2)$ identities are used to put the dimension-6 contributions in a standard form. 
Note that Eq.~\eqref{eq:newl6} contains dimension-$8$ interactions in addition to those of the dimension-$8$ operator of $\mathcal{L}_8$. 
Repeated use of the SM equations of motion on the dimension-$8$ term, $\delta \mathcal{L}_{8a}$,  yields the second line of the expression for $\delta L_{8a}$ in Eq.~\eqref{eq:newl6}.
We have not simplified the dimension-8 contributions with derivatives on the Higgs doublet, since it is straightforward to use \textsc{FeynRules}~\cite{Degrande_2012,Alloul_2014} to determine the needed interactions for a given application. 
The operators are defined in Tab.~\ref{tab:opdef} using the bases of Refs.~\cite{Grzadkowski_2010,Murphy:2020ab}.

We simplify the dimension-$8$ operator of Eq.~\eqref{eq:theory} to extract the term contributing to the top quark Yukawa interaction,
\begin{eqnarray}
\mathcal{L}_8& \rightarrow & {\lambda_t^3\lambda_T^2\over 2m_{\mathcal{T}}^4}(H^\dagger H)^2
\biggl({\overline{\psi}}_L{\tilde H} t_R + \textrm{h.c.} \biggr)+
{\rm terms ~with~ derivatives ~on}~ {\tilde{H}}~{\rm or}~{\tilde{H}}^\dagger
\nonumber \\ &\equiv & \biggl({\lambda_t^3\lambda_T^2\over   2m_{\mathcal{T}}^4}(H^\dagger H)^2
O_{quH^{5}}^{(8)} + \textrm{h.c.} \biggr) + \delta \mathcal{L}_{8b}\, ,
\end{eqnarray}
where the complete expression for $\delta \mathcal{L}_{8b}$ can be found in the supplemental material. The contribution to $\delta \mathcal{L}_{8b}$ that is proportional to the strong coupling, $g_s$, is given in Appendix~\ref{sec:dim8g} and the momentum dependence of the dimension-8 operators is clearly seen. 

The complete SMEFT Lagrangian generated from the TVLQ model to dimension-$8$ involving the top quark written in terms of the Lagrangian parameters is,
\begin{eqnarray}
\mathcal{L}&\equiv & \mathcal{L}_4^\prime + \mathcal{L}_6^\prime + \mathcal{L}_8^\prime \nonumber \\
\mathcal{L}_4^\prime &=& {\overline \psi}_L(i\slashed{D}) \psi_L+{\overline t_R(i\slashed{D})t_R} -(\lambda_t {\overline \psi}_L{\tilde H} t_R + \textrm{h.c.})
\nonumber \\ 
\mathcal{L}_6^\prime
&=&\biggl( { \lambda_t\lambda_T^2\over 2 m_{\mathcal{T}}^2} O_{tH}^{(6)}+ \textrm{h.c.} \biggr)+{\lambda_T^2\over 4 m_{\mathcal{T}}^2}
\biggl(O_{Ht}^{1, (6)}-O_{Ht}^{3,(6)}\biggr)\nonumber \\ 
\mathcal{L}_8^\prime & = & {\lambda_t\lambda_T^2\over 8m_{\mathcal{T}}^4}(4\lambda_t^2-3\lambda_T^2) O_{quH^{5}} ^{(8)}+\delta \mathcal{L}_{8b}+ \textrm{h.c.}
\end{eqnarray}
We note that changing the input parameters from ($\lambda_t$, $\lambda_T$, $m_{\mathcal{T}}$)  to ($m_t$, $M_T$, $s_L$) using Eq.~\eqref{eq:rels}
re-arranges the counting in terms of inverse powers of the heavy mass~\cite{Brehmer:2015rna}.   Ref.~\cite{Brehmer:2015rna} argues that replacing 
the Lagrangian mass, $m_{\mathcal{T}}$, with the physical mass, $M_T$, improves the agreement
between the SMEFT predications and those of the corresponding UV complete model in many cases.  A similiar effect is found in the EFT limit
of the 2HDM~\cite{Egana-Ugrinovic:2015vgy,B_lusca_Ma_to_2017}.

The terms contributing to the SMEFT relationship between the top mass and  Higgs top Yukawa coupling are,
\begin{eqnarray}
\mathcal{L} &\sim& -\lambda_t {\overline{\psi}}_L {\tilde {H}}t_R +{C_{tH}^{(6)}\over m_{\mathcal{T}}^2}
 O_{tH}^{(6)}+
{C_{quH^{5}}^{(8)} \over m_{\mathcal {T}}^4} 
O_{quH^{5}}^{(8)} + \textrm{h.c.} 
\label{eq:smeftyuk} 
\end{eqnarray}
with,
\begin{eqnarray}
 C_{tH}^{(6)}={\lambda_t\lambda_T^2\over 2} \, ,
\qquad  C_{quH^{5}}^{(8)}={\lambda_t\lambda_T^2\over 8} (4\lambda_t^2-3\lambda_T^2)
\, . 
\label{eq:yukops}
\end{eqnarray}
It is interesting to study the behaviors of $C_{tH}^{(6)}$ and $C_{quH^5}^{(8)}$ using the relationships of Eq.~\eqref{eq:exact} and expanding in powers of ${1 / m_{\mathcal{T}}}$ keeping the top quark mass fixed to its physical value.\footnote{Note that we are free to take a combination of three Lagrangian and/or physical parameters as inputs.}  Note that keeping the top quark mass fixed rearranges the counting, as does alternatively using $s_L$ and $M_T$ as inputs.    
To ${\cal{O}}({1/ m_{\mathcal{T}}^4})$,
\begin{eqnarray}
\lambda_t&=& {\sqrt{2}m_t\over  v}\biggl\{1+{\lambda_T^2v^2\over 4 m_{\mathcal{T}}^2} 
+{\lambda_T^2v^2 m_t^2 \over 4 m_{\mathcal{T}}^4}
-{\lambda_T^4 v^4\over 32 m_{\mathcal{T}}^4}\biggr\} 
\nonumber \\ 
&\rightarrow&  {\sqrt{2}m_t\over c_L v}\biggl\{ 1-s_L^2{m_t^2\over m_{\mathcal{T}}^2}-s_L^4 {m_t^4\over  4m_{\mathcal{T}}^4}\biggr\}\nonumber \\ 
{\overline {C}}_6&\equiv & {C_{tH}\over m_{\mathcal{T}}^2}\nonumber \\
&\rightarrow& {2m_t s_L^2\over \sqrt{2} c_L^3 v^3}\biggl(1-{(s_L^2+4)m_t^2\over 2m_{\mathcal{T}}^2}
-{(s_L^4-8s_L^2-8)m_t^4 \over 8 m_{\mathcal{T}}^4}\biggr)\nonumber \\
{\overline {C}}_8&\equiv & {C_{quH^5}\over m_{\mathcal{T}}^4}\nonumber \\
&\rightarrow& {m_t\over\sqrt{2}}{s_L^2\over c_L^5 v^5} \biggl\{ -3s_L^2 +{m_t^2\over 2 m_{\mathcal{T}}^2}(3s_L^4+24 s_L^2+8)
 +{m_t^4\over 8 m_{\mathcal{T}}^4}(3s_L^6-48s_L^4-192 s_L^2-64)\biggr\}\, .
 \label{eq:coefexp}
\end{eqnarray}
The naive scalings, ${\overline {C}}_6\sim {1/ m_{\mathcal{T}}^2}$  and ${\overline {C}}_8\sim {1 / m_{\mathcal{T}}^4}$ are modified 
by terms of ${\cal{O}}(s_L^2 x)$ when using the physical parameters.

Expanding Eq.~\eqref{eq:smeftyuk} to linear order in the Higgs field, we define
the top Yukawa, $Y_t^{(8)}$, as usual 
\begin{equation}
\mathcal{L} \sim -{Y_t^{(8)}\over\sqrt{2}} {\overline  t}_Lt_R h + \textrm{h.c.}
\end{equation}
where the superscript $8$ denotes the inclusion of the dimension-$8$ contributions.
We initially fix $m_t$ (the physical top quark mass), $\lambda_T$ and $m_{\mathcal{T}}$, 
\begin{eqnarray}
Y_t^{(8)}&=&{\lambda_t}-{3v^2\over 2}{\overline {C}}_6-{5v^4\over  4}{\overline {C}}_8\nonumber \\
&=&{m_t\sqrt{2}\over v}\biggl\{ 1-{\lambda_T^2 v^2\over 2 m_{\mathcal{T}}^2}-{m_t^2 v^2 \lambda_T^2\over m_{\mathcal{T}}^4}
+{\lambda_T^4v^4\over 4 m_{\mathcal{T}}^4}\biggr\}\, .
\label{eq:ytm}
\end{eqnarray}
Retaining only the dimension-$6$ terms in the Lagrangian,
\begin{eqnarray}
Y_t^{(6)}&=&{\lambda_t}-{3v^2\over 2}{\overline {C}}_6\nonumber\\
&=&{m_t \sqrt{2} \over v}\biggl\{ 1-{ \lambda_T^2 v^2\over 2m_{\mathcal{T}}^2}\biggr\}+{\cal{O}}\biggl({1\over m_{\mathcal{T}}^4}\biggr)
\, .
\label{eq:yt6def}
\end{eqnarray}

In Eqs.~\eqref{eq:ytm} and~\eqref{eq:yt6def}, the SM is recovered in the ${\lambda_T^2 / m_{\mathcal{T}}^2}\rightarrow 0 $ limit, which corresponds to the $s_L\rightarrow 0$ limit.  The choice to use $m_t$ as an input introduces
terms of ${\cal{O}}\big( {\lambda_T^4 / m_{\mathcal{T}}^4}\big) $ in $Y_t^{(6)}$ due to the interdependence of the parameters. 

\begin{figure}
\centering
\includegraphics[width=0.47\textwidth]{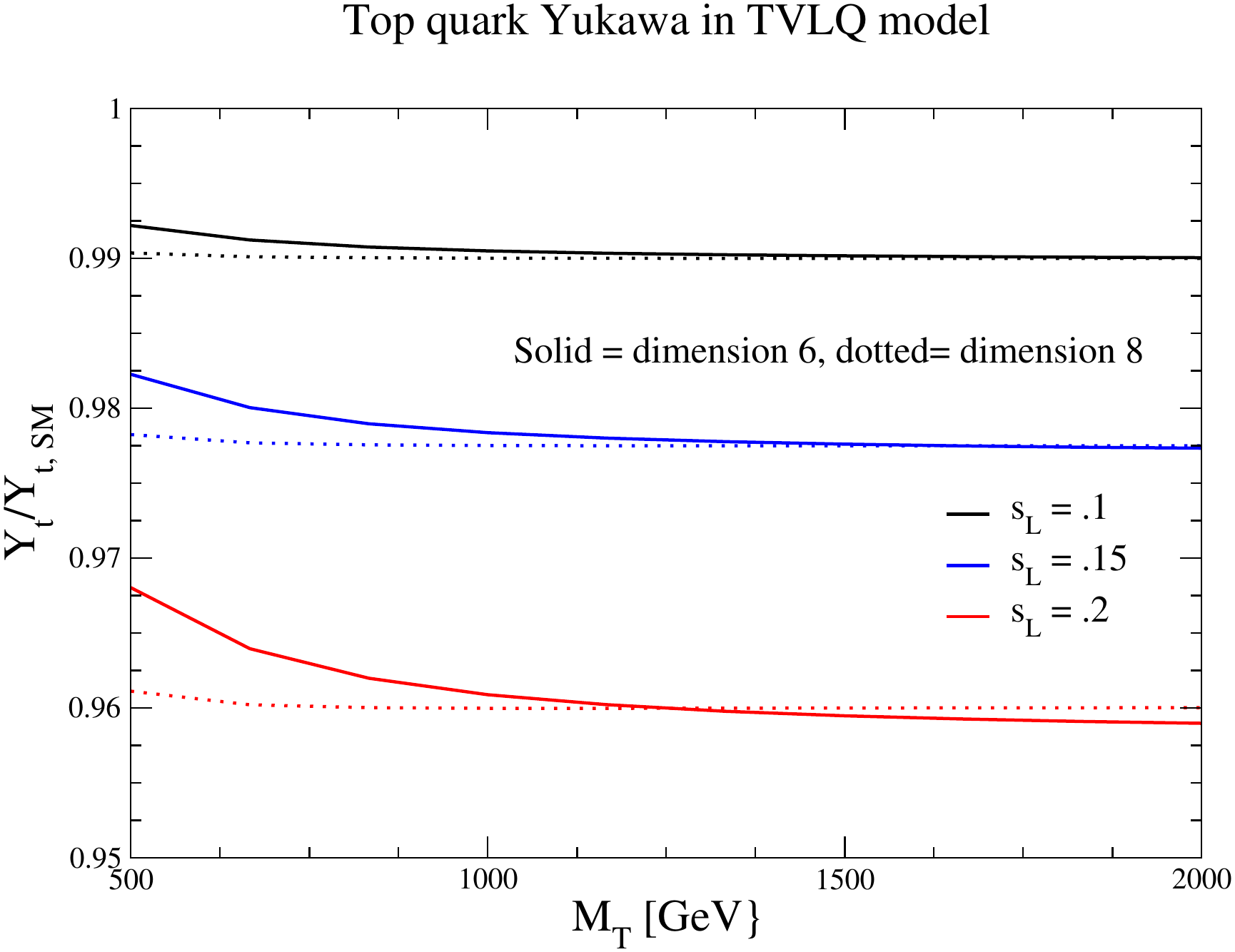}
\caption{Top quark Yukawa coupling  normalized to the SM top quark Yukawa when dimension-$8$ contributions are included ($Y_t^{(8)}$) 
 and when only dimension-$6$ terms are included ($Y_t^{(6)}$).}
\label{fig:yukrat}\
\end{figure}

In the small $s_L$ limit,
\begin{eqnarray}
Y_t^{(8)} &\rightarrow &{m_t\sqrt{2}\over v}\biggl\{ 1-s_L^2 +{3s_L^2 m_t^4\over m_{\mathcal{T}}^4}\biggr\}
\nonumber \\ 
Y_t^{(6)}&\rightarrow &{m_t\sqrt{2} \over v}\biggl\{ 1-s_L^2+{5s_L^2 m_t^2\over 2m_{\mathcal{T}}^2}-{3 s_L^2m_t^4\over 2 m_{\mathcal{T}}^4}\biggr\} \, . 
\end{eqnarray}
We see that the SM limit is only recovered in the $s_L\rightarrow 0$ limit,
 consistent with the decoupling discussion in the previous section.  
Fig.~\ref{fig:yukrat}  shows  the effect of including the dimension-$8$ terms on the top quark Yukawa coupling and we see that it is typically less than a few percent 
for $500~\textrm{GeV} < m_{\mathcal{T}}< 1~\textrm{TeV}$.

\section{Phenomenology}
\label{sec:tthpheno}

We are now in a position to investigate the numerical effects of including the dimension-8 terms in the SMEFT analysis of the TVLQ and in the comparison between
SMEFT and the UV complete model.  As an example of the possible impact of the dimension-8 contributions, we 
consider $t \bar{t} h$ production at the $13~\textrm{TeV}$ LHC\footnote{The TVLQ model contributes to 
gluon fusion at one-loop, but a consistent inclusion of the dimension-8 contributions would require
the double insertion of the dimension-6 contributions. The contribution
to $gg\rightarrow h$ from the TVLQ is suppressed by $s_L^2$ and is numerically
small\cite{Dawson:2012di}.}.
In addition to the SM cross section, $\dd\sigma_{\textrm{SM}}$, we consider various SMEFT expansions:
\begin{eqnarray}
\dd \sigma_{\textrm{int}}  &\sim &\int (\dd {\rm PS})\biggl\{ | \mathcal{A}_{\textrm{SM}}| ^2 +{2\over \Lambda^2} 
\textrm{Re}\biggl(\sum C_i^{(6)} \mathcal{A}_i ^{(6)} \mathcal{A}_{\textrm{SM}}^*\biggr)
\biggr\}
\nonumber \\
\dd \sigma_{6} &\sim &\int (\dd {\rm PS})\biggl\{
 | \mathcal{A}_{\textrm{SM}}| ^2 +{2\over \Lambda^2} 
\textrm{Re}\biggl(\sum C_i^{(6)} \mathcal{A}_i ^{(6)} \mathcal{A}_{\textrm{SM}}^*\biggr)
+{1\over \Lambda^4}
\textrm{Re}\biggl(\sum_{i, j}
C_i^{(6)}C_j^{(6)*} \mathcal{A}_i^{(6)} \mathcal{A}_j^{(6)~*}\biggr)
\biggr\}
\nonumber \\
\dd \sigma_{8} &\sim &\int (\dd {\rm PS})\biggl\{
 | \mathcal{A}_{\textrm{SM}}| ^2 +{2\over \Lambda^2} 
\textrm{Re}\biggl(\sum C_i^{(6)} \mathcal{A}_i ^{(6)} \mathcal{A}_{\textrm{SM}}^*\biggr)
+{1\over \Lambda^4}
\textrm{Re}\biggl(\sum_{i, j}
C_i^{(6)}C_j^{(6)*} \mathcal{A}_i^{(6)} \mathcal{A}_j^{(6)~*}\biggr)
\nonumber \\
&&\quad +{2\over \Lambda^4} \textrm{Re} \biggl(\sum_i C_i^{(8)} \mathcal{A}_i^{(8)} \mathcal{A}_{\textrm{SM}}^*\biggr)\biggr\}\, .
\label{eq:expand} 
\end{eqnarray}
In particular, $\dd \sigma_{6}$ and $\dd\sigma_{8}$ are of the same order in  $1/\Lambda^4$ and the difference between the two is a measure of the importance of the dimension-8 terms.  In our numerical studies, we will always take $\Lambda=m_{\mathcal{T}}$.

The rescaling of the top Yukawa coupling at dimension-8 will give only  a small difference from the dimension-6 result as demonstrated in Fig.~\ref{fig:yukrat}. However,
the dimension-8 terms introduce a momentum dependence into the $t \bar{t}h$ and $t\bar{t}hg$ vertices, as well as the $tbW$ and $tbWh$ vertices. The Feynman rules relevant for the $t {\bar{t}}h$ process are given in Appendix~{\ref{sec:dim8g}. 
Note that, since there is never more than one covariant derivative operating on the top quark at dimension-$8$, the TVLQ model only generates new operators with a single gluon field.
We use \textsc{FeynRules}~\cite{Christensen_2009,Alloul_2014} to generate the Feynman rules including the dimension-8 terms and use the resulting UFO~\cite{Degrande_2012} file with \textsc{MadGraph5}  and the default dynamical scale choice  to generate events. 
For all our simulations, we set $m_t = 172\,\textrm{GeV}$, $m_h = 125\,\textrm{GeV}$, $m_Z = 91.1876\,\textrm{GeV}$, $G_F = 1.16637 \times 10^{-5}$, and $\alpha = 1/127.9$, so that $m_W$ is computed to be $79.82436\,\textrm{GeV}$ at tree level in
the \textsc{MadGraph5} code.  We use the NNPDF23 LO PDF set with $\alpha_s=0.119$.
The complete set of interactions can be found using the \textsc{FeynRules} module contained in the supplemental materials.

\begin{figure}[ht]
\centering
\includegraphics[width=0.3\textwidth]{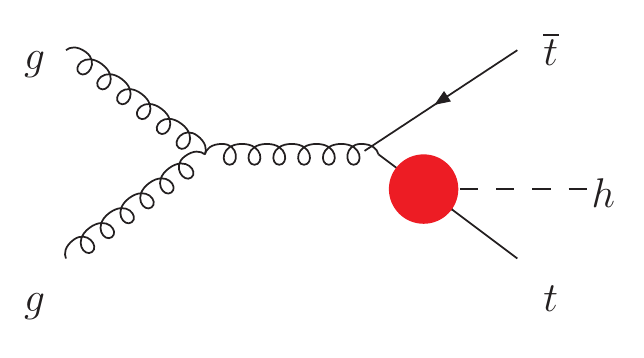}
\includegraphics[width=0.3\textwidth]{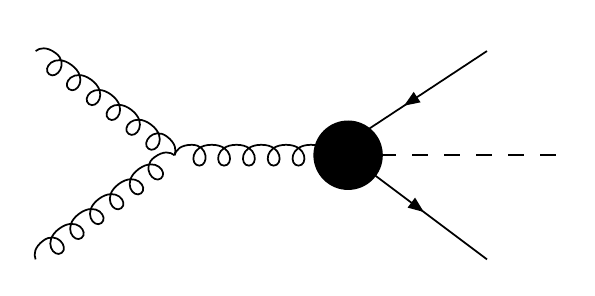}
\includegraphics[width=0.26\textwidth]{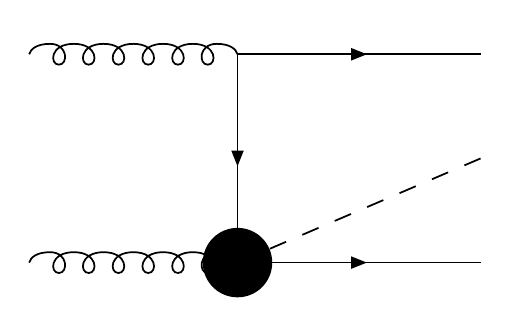}
\caption{Sample Feynman diagrams contributing to $gg\rightarrow t \bar{t}h$ including dimension-$8$ operators. The blobs represent the effects of the dimension-$8$ operators for the center and right-handed diagrams, while the left-most diagram (red blob) receives contributions from both dimension-$6$ and dimension-$8$ operators. 
Similar diagrams with $q\bar{q}$ initial states are not shown, but included in our computations.
}
\label{fig:fdiags1}
\end{figure}

We begin by considering the $t \bar{t} h$ process without decays. 
Some sample diagrams are shown in Fig.~\ref{fig:fdiags1}.
There are two effects from the higher dimension operators. 
The first is the rescaling of the $t\bar{t}h$ Yukawa interaction. This does not  lead to any momentum-dependent effects in the $gg \to t\bar{t}h$ process, but due to the small contributions from the $q\bar{q} \to t\bar{t} h$ sub-process, where the $h$ couples to an intermediate $Z$ boson, which are not rescaled by the top Yukawa, there are very small kinematic effects in the total cross section at the $\lesssim 1\%$ level.
The second effect, which first arises at dimension-$8$, is interactions that are enhanced by an
energy factor, $\sim s$ (with $s$ the partonic center of mass energy), relative to the SM contributions, both in the $t\bar{t} h$ and $t\bar{t}hg$ effective vertices.
However, these $s$-enhanced contributions are proportional to a difference in the projection operators, $(P_R - P_L)$ (c.f. Eq.~\eqref{eq:frg}), and the enhancement is therefore averaged out in the helicity-blind production of on shell tops from QCD production.
The resulting distributions for $t\bar{t} h$ production without decays are therefore essentially flat in various kinematic observables, and roughly consistent with an overall rescaling of the cross section by the modified top Yukawas in Eqs.~\eqref{eq:ytm} and \eqref{eq:yt6def}.

\begin{figure}[ht]
\centering
\includegraphics[width=.3\textwidth]{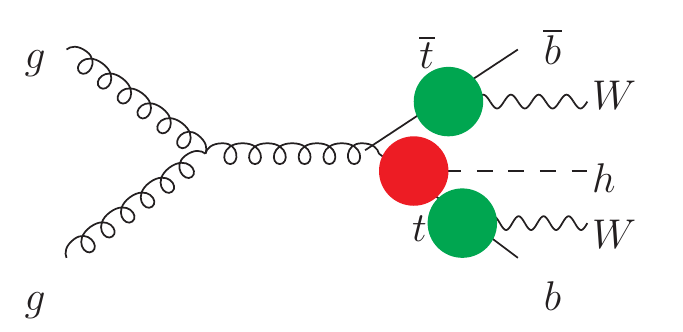}
\includegraphics[width=0.3\textwidth]{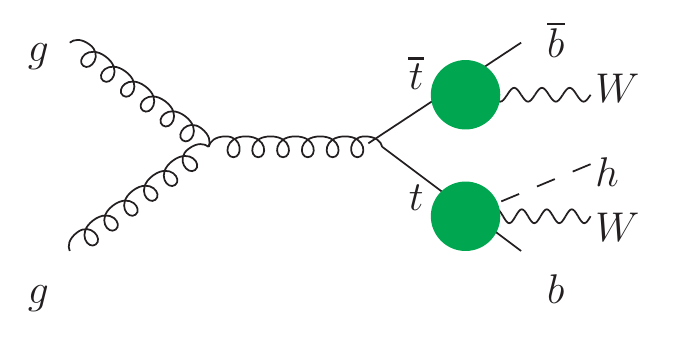}
\caption{Sample diagrams contributing to $gg \rightarrow  b {\overline{b}} W^+W^-h$. The green blob represents the insertion of dimension-$6$ operators, while the red blob represents both dimension-$6$ and dimension-$8$ operators. Similar diagrams with $q\bar{q}$ initial states are not shown, but included in our computations.
}
\label{fig:fdiags2}
\end{figure}

\begin{figure}[h]
\centering
\includegraphics[width=0.65\linewidth]{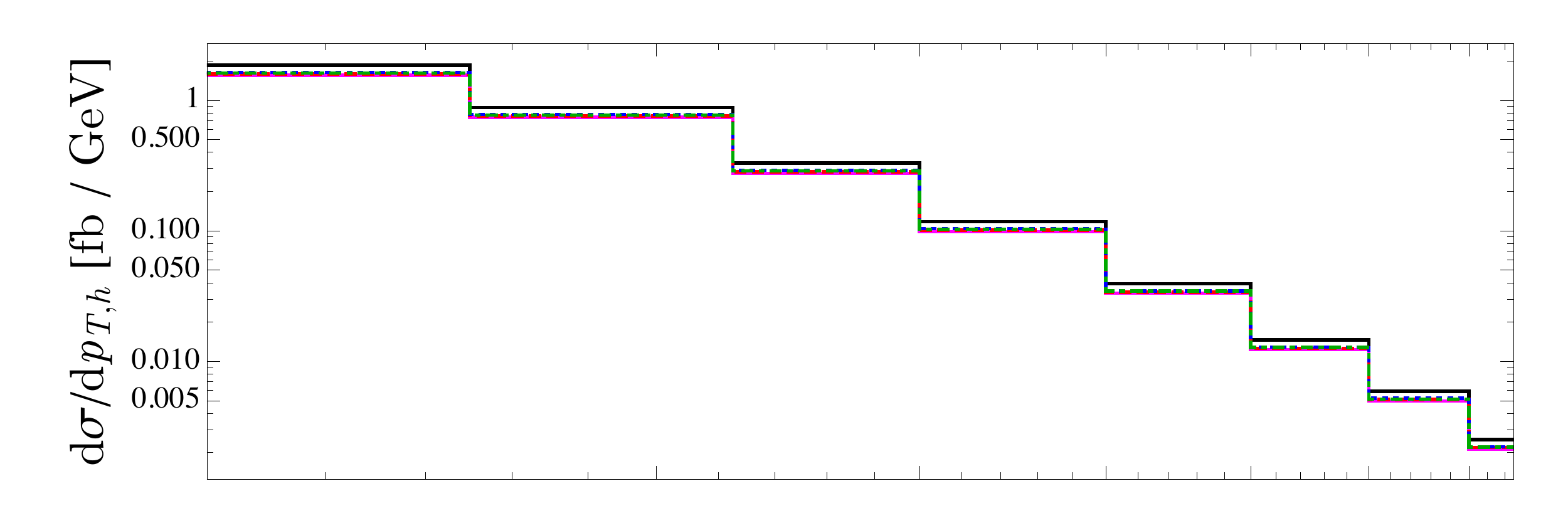}
\vskip -0.3 cm
\includegraphics[width=0.65\linewidth]{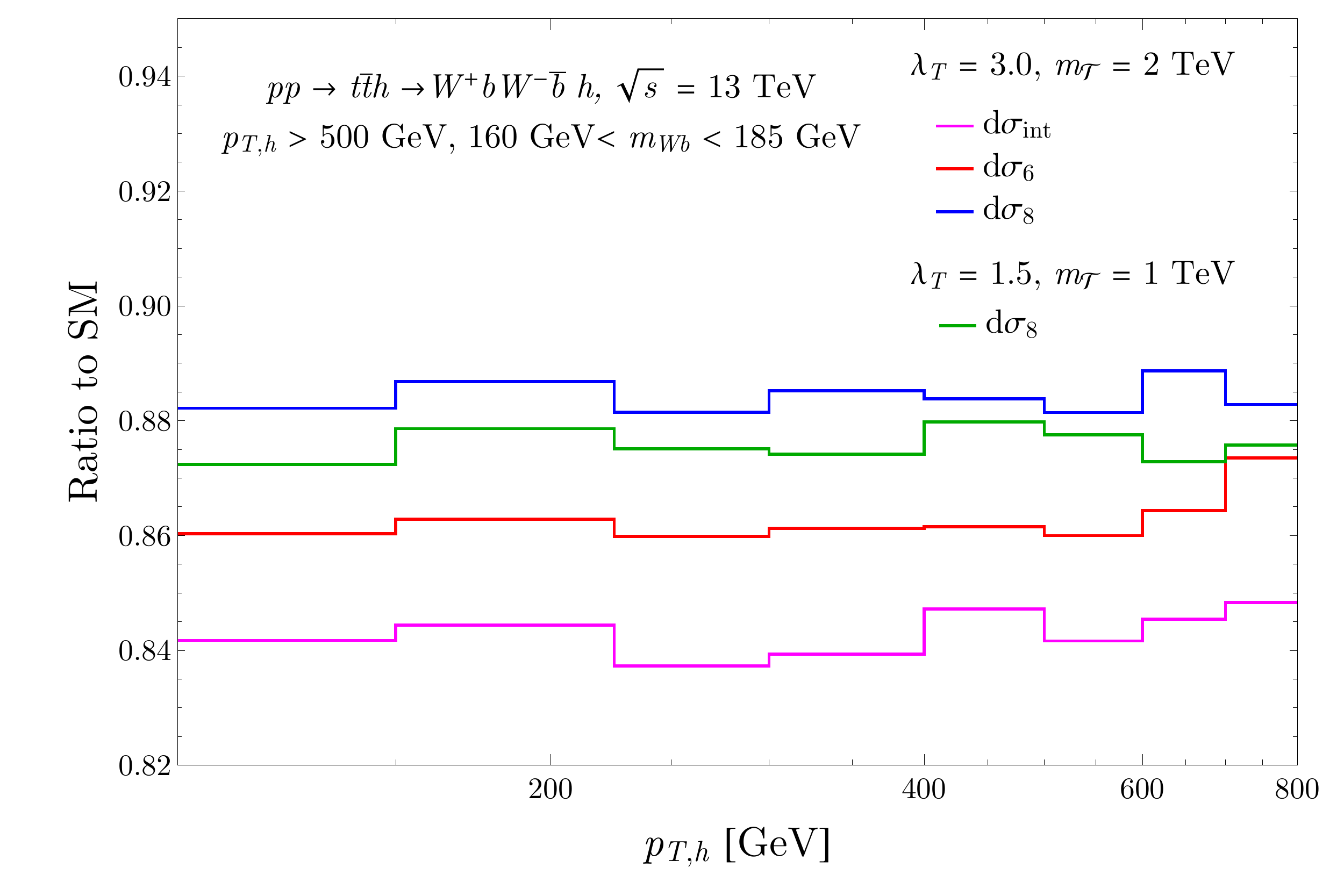}
\vskip -0.5cm
\caption{
The distribution for $W^+W^- b\bar{b} h$ production with intermediate top quarks in bins of the Higgs $p_{T,h}$.
The top panel shows the distribution for the SM (black), and for the TVLQ point with $\lambda_{T} = 3.0$, $m_{\cal T} = 2\,\textrm{TeV}$, matched to various orders in the SMEFT, as in Eq.~\eqref{eq:expand}, with the magenta, red and blue curves representing $\dd\sigma_{\textrm{int}}$, $\dd\sigma_{6}$ and $\dd\sigma_{8}$, respectively.
The results for $\dd\sigma_{8}$ with $\lambda_T = 1.5$ and $m_{\cal T} = 1\,\textrm{TeV}$ are shown in green.
}
\label{fig:tth_dist_decayed}
\end{figure}

We next consider $t\bar{t}h$ production with the tops decayed to the final state $b\bar{b} W^+ W^- h$.
We generate events in this final state from all tree level diagrams including intermediate top quarks to exclude pure electroweak production of $W$ and $b$ pairs.
This includes contributions from a number of diagrams which cannot be factorized into $t\bar{t} h$ production times decay. One example of such a diagram is shown in the right-hand side of Fig.~\ref{fig:fdiags2}. There are also contributions that are not proportional to the top Yukawa coupling, where the Higgs instead couples to the $W$ bosons or bottom quarks.

We compute the cross section for $p p \to b\bar{b} W^+ W^- h$ with intermediate top quarks using our \textsc{FeynRules} implementation, and plot the result in bins of $p_{T,h}$ in Fig.~\ref{fig:tth_dist_decayed}.
We show results both for the SM, and with the SMEFT matched with $\lambda_T = 3.0$ and $m_{\mathcal{T}} = 2\,\textrm{TeV}$, corresponding to a mixing angle $\sin\theta_L \sim 0.25$. We note that such large values of the mixing angle are excluded by fits to the oblique parameters (see Fig.~\ref{fig:stu}) --- we choose such a large point to make the kinematic effects that arise at different orders in the SMEFT expansion clear.
To focus on the effects on $t\bar{t} h$ production, we impose a cut on the $W$ boson and $b$ quark system, requiring it to be near the top quark mass shell: $160\,\textrm{GeV} < m_{Wb} < 185\,\textrm{GeV}$.
We utilize the charge information of the $W$ and $b$ particles in performing this cut, assuming that they can be properly assigned to the correct top quark in a true experimental analysis, e.g., if they are all identified in a single large-radius top jet.

Including the full $b\bar{b}W^+W^- h$ final state changes the expectations from $t\bar{t} h$ production without decays significantly.
The diagrams where the Higgs is coupled to a $W$ boson or $b$ quark are not proportional to the top Yukawa, and therefore are not rescaled by the corrections to the top Yukawa as the bulk of the cross section is in the un-decayed case.
This leads to a growth in the cross section for large $p_{T,h}$ even at the dimension-6 level, and a change in the overall rate that is significantly different from a naive rescaling.
At dimension-8, there are non-factorizable contributions with $t h W b$ vertices, which have one fewer propagator than the SM-like diagrams, and as a result, $s$-enhanced effects relative to the Standard Model.
Finally, since the tops decay via their $SU(2)_L$ interactions, the effective operators proportional to $(P_R - P_L)$ discussed above will no longer be averaged out, and can therefore lead to additional effects at high $p_{T,h}$ as well.
All of these effects in the amplitudes compete, and interfere with one another.

The resulting effects in Fig.~\ref{fig:tth_dist_decayed} show that the kinematic effects apparent at dimension-6 are nearly washed out at dimension-8, and the distribution is almost flat. 
We emphasize that, while the overall distribution is roughly flat in $p_{T,h}$, due to a combination  of different effects that arise at different orders in the EFT expansion, the overall rate is different than that expected by rescaling the SM cross section by the modified top Yukawa.
Note also that the size of the contributions from the dimension-8 operators are similar to the size of the dimension-6 squared terms relative to the interference contribution alone.

We also include in Fig.~\ref{fig:tth_dist_decayed} a curve in green, showing the results for matching up to dimension-8 with $\lambda_T = 1.5$ and $m_{\mathcal{T}} = 1\,\textrm{TeV}$. These values are chosen such that the effects at dimension-6 in the SMEFT, which all scale as $\lambda_T^2 / m_{\mathcal{T}}^2$, are precisely the same as for $\lambda_T = 3.0$, $m_{\mathcal{T}} = 2\,\textrm{TeV}$. At dimension-8, however, there are effects that break this scaling (c.f. Eqs.~\eqref{eq:ytm} and \eqref{eq:frg} -- \eqref{eq:frtbwh}), and we see that indeed, the dimension-8 curve in green is different than the curve in blue.

\begin{figure}
\includegraphics[width=0.48\linewidth]{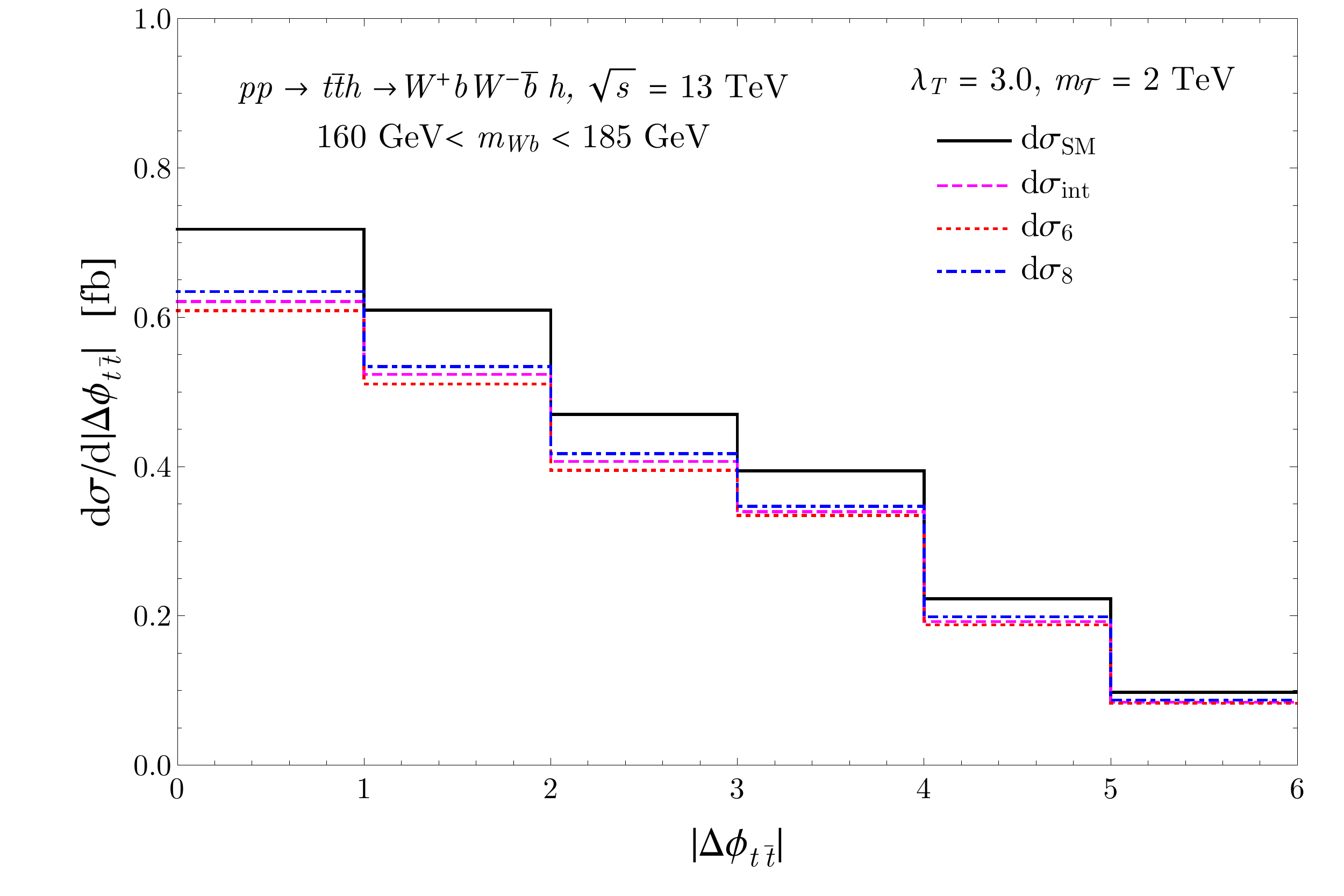}
~
\includegraphics[width=0.48\linewidth]{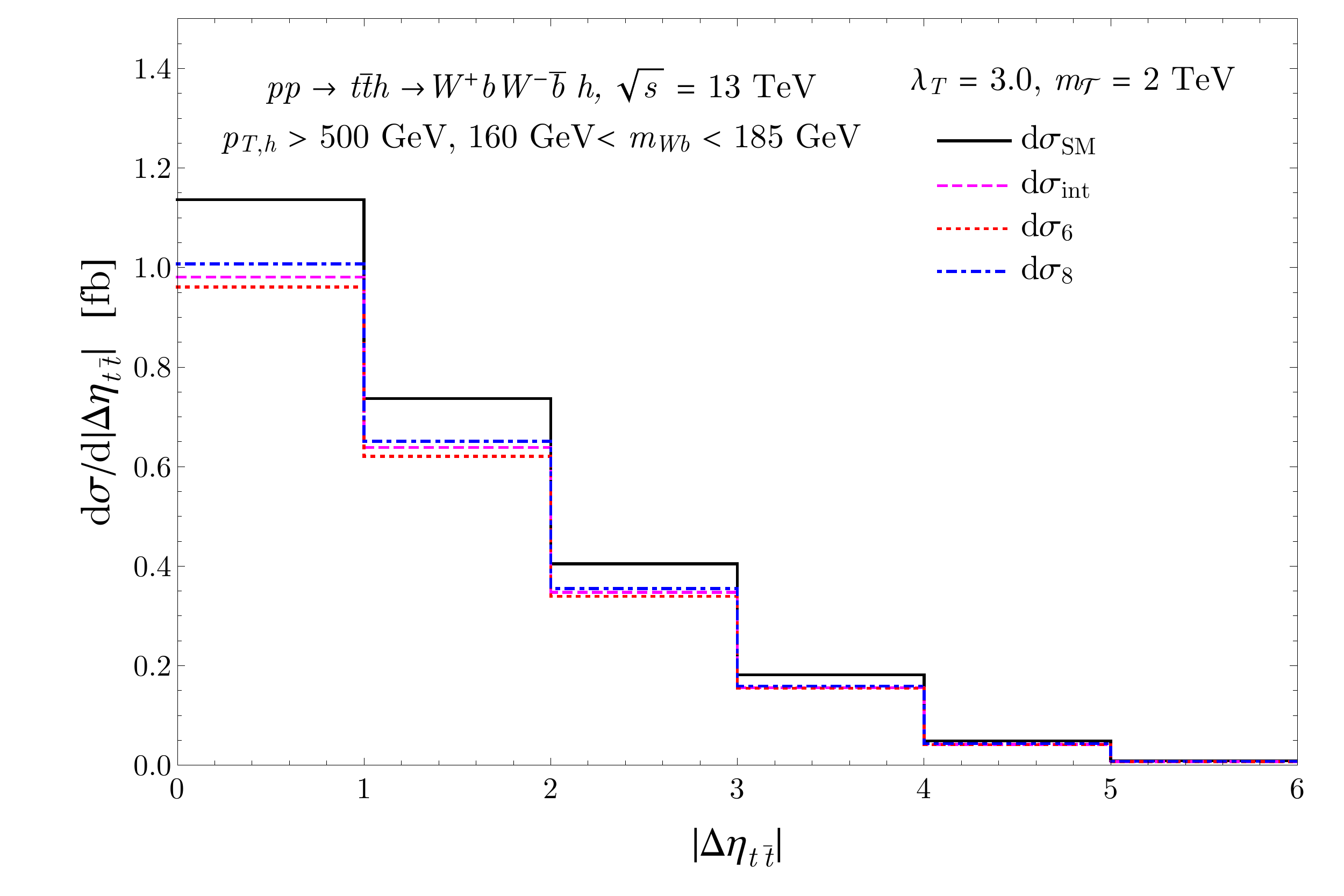}
\vskip -0.25cm
\caption{The distribution for $W^+W^-b\bar{b}h$ production with intermediate top quarks in bins of $|\Delta\phi_{t\bar{t}}|$ (left) and $|\Delta\eta_{t\bar{t}}|$ (right), for events with $p_{T,h} > 500\,\textrm{GeV}$, matched at various orders in the SMEFT expansion for $\lambda_T = 3.0$, $m_{\mathcal{T}} = 2\,\textrm{TeV}$. The black solid, magenta dashed, red dotted, and blue dot-dashed curves show the results for $\dd\sigma_{\textrm{SM}}$, $\dd\sigma_{\textrm{int}}$, $\dd\sigma_{6}$ and $\dd\sigma_8$, respectively.}
\label{fig:tth_dists_angular}
\end{figure}

In Fig.~\ref{fig:tth_dists_angular}, we show the distributions for $t\bar{t}h$ production including the full $b\bar{b}W^+W^- h$ final state in bins of $|\Delta\phi_{t\bar{t}}|$ and $|\Delta \eta_{t\bar{t}}|$, after placing a cut on the Higgs $p_{T,h} > 500\,\textrm{GeV}$.
We see there are no kinematic effects  in these distributions at any order in the SMEFT expansion, other than the rescaling consistent with the results in Fig.~\ref{fig:tth_dist_decayed}.

Finally, we comment on the size of the dimension-8 effects for parameters that are not experimentally excluded.
We take $\lambda_T = 1.5$, $m_{\mathcal{T}} = 2\,\textrm{TeV}$, corresponding to a mixing angle $\sin\theta_L = 0.13$, which is near the edge of the region allowed by the oblique parameter fits shown in Fig.~\ref{fig:stu}.
For these parameters, the effects of the dimension-8 terms included in $\dd\sigma_8$ are very small: $\lesssim \mathcal{O}(1\%)$ of the SM cross section. The effects of the $\mathcal{O}(1/\Lambda^4)$ terms in $\dd\sigma_6$ are of similar size.
There are small kinematic effects, but the total rate is quite similar to what one expects from a naive rescaling of the cross section by $(Y_t^{(6)})^2$. 
We conclude that the effects of the $\mathcal{O}(1/\Lambda^4)$ terms are too small to affect constraints  on the TVLQ model at the LHC from $tth$ production.

\section{Discussion}
\label{sec:discussion}
We have implemented the complete dimension-8 set of operators contributing to the tree level process, $p p
\rightarrow t {\overline t}h$, in a model with a charge-$2/3$ vector-like quark.  When the decays of the top quark are not included, the results are almost entirely given by the rescaling of the top quark Yukawa coupling. The decays of the top quark introduce a momentum dependence due primarily to the presence of non-factorizable $tbWh$ vertices.  These effects create a difference of 
less than ${\cal O}(2\%)$  at high $p_{T,h} $ between the square
of the dimension-6 contributions and the result with the dimension-8 contributions included.

The example we have considered is particularly simple, since the input parameters are not rescaled at tree level to 
dimension-8.  It would be of interest to consider the effects of a more complicated model
which generates tree level rescaling of the input parameters at dimension-8.  The results of Refs.
\cite{Corbett:2021iob,Corbett:2021eux} suggest that the dimension-8 contributions may play a more significant role in
such scenarios.

The UFO and \textsc{FeynRules} model files used to generate the TVLQ dimension-8 effects are 
included as supplemental material.

\section*{Acknowledgments}

SD  and MS are supported by the United States Department of Energy under Grant Contract DE- SC0012704. 
SH is supported in part by the DOE Grant DE-SC0013607, and in part by the Alfred P. Sloan Foundation Grant No. G-2019-12504. Digital data is contained in the supplemental material submitted with this paper.

\appendix

\section{Dimension-8 Interactions}
\label{sec:dim8g}

The following terms in the tree-level dimension-$8$ Lagrangian, $\mathcal{L}_8^\prime$,  contain non-SM gluon couplings:
\begin{eqnarray}
\mathcal{L}_{8,g} &=& i \frac{\lambda_T^2}{m_{\mathcal{T}}^4} (\bar{\psi}_L D^\mu \tilde{H}) \gamma_\mu (D^\nu \tilde{H}^\dagger D_\nu \psi_L) 
        + i \frac{\lambda_T^2}{m_{\mathcal{T}}^4} (\bar{\psi}_L \tilde{H}) \gamma_\mu (D^\mu D^\nu \tilde{H}^\dagger D_\nu \psi_L) \nonumber\\
        &-& i \frac{\lambda_T^2}{m_{\mathcal{T}}^4} (\bar{\psi}_L \tilde{H}) \gamma_\mu (D^\nu D^\mu \tilde{H}^\dagger D_\nu \psi_L) 
        - \frac{\lambda_t\lambda_T^2}{m_{\mathcal{T}}^4} (H^\dagger H) \bar{t}_R (D^\mu \tilde{H}^\dagger D_\mu \psi_L) 
        + \textrm{h.c.} ,
\end{eqnarray}
where indices are contracted implicitly such that terms in parentheses are $SU(2)$ singlets.
The $t {\overline {t}}h$ interactions that are needed for the tree level process are
(with all momenta outgoing), 
\begin{eqnarray}
t(p_1){\overline t}(p_2) h(p_h)&=& -i {Y_t^{(8)}\over\sqrt{2}} + i{\lambda_T^2 v m_t\over 4 m_{\mathcal{T}}^4} (p_1-p_2)\cdot p_h(P_R-P_L)\nonumber \\
t(p_1){\overline t}(p_2) h(p_h)g^{A \mu} (p_g)&=& i g_s {\lambda_T^2 v m_t\over 2 m_{\mathcal{T}}^4}p_h^\mu T^A(P_R-P_L)
\label{eq:frg}
\end{eqnarray}
where $P_{L,R}\equiv {1\over 2}(1\mp \gamma_5)$ and $Y_t^{(8)}$ is as given in Eq.~\eqref{eq:ytm}. 

The following are the electroweak couplings of the top quark expanded to dimension eight that
occur in the $t {\overline {t}}h, ~t\rightarrow Wb$ process:
\begin{eqnarray}
b(p_b){\overline t}(p_t) W^{+\mu}(p_W) &=&
 i \frac{g_W}{\sqrt{2}} \gamma^\mu P_L 
 \biggl[1-  \frac{ v^2 \lambda_T^2}{4  {m_\mathcal{T}}^2}  +
   \frac{v^2 \lambda_T^2}{32  {m_\mathcal{T}}^4} 
   \biggl(3 v^2 \lambda_T^2 + 8 (m_t^2 - p_t^2+p_b^2)
   \biggr)  \biggr]
   \nonumber \\ 
   && - i \frac{g_W v^2 \lambda_T^2 }{2 \sqrt{2} {m_\mathcal{T}}^4} p_b^\mu P_R 
   \biggl[ m_t- \slashed{ p_W}\biggr] \label{eq:frtbw}
\end{eqnarray}
\begin{eqnarray}
b(p_b){\overline t}(p_t) W^{+\mu}(p_W) h(p_h) &=& 
- \frac{i v g_W \lambda_T^2 \gamma^{\mu} P_L}{2 \sqrt{2} {m_\mathcal{T}}^2}   \biggl[1
+ \frac{  p_{h }\cdot  p_{t}}{{m_\mathcal{T}}^2} 
+ \frac{ 2 p_{b }\cdot p_{W} } {{m_\mathcal{T}}^2}  + \frac{ p_h^2 }{2 {m_\mathcal{T}}^2}  +\frac{ p_W^2 } {{m_\mathcal{T}}^2} 
+ \frac{p_h\cdot p_W}{{m_\mathcal{T}}^2} 
\nonumber \\ &&
- \frac{3  v^2  \lambda_T^2 }{4  {m_\mathcal{T}}^2} 
- \frac{2 m_t^2 }{ {m_\mathcal{T}}^2}
- \frac{\slashed{p_h}\slashed {p_W}}{2  {m_\mathcal{T}}^2}
-{m_t \slashed{p_h}\over 2 m_{\mathcal{T}}^2}
\biggr]
\nonumber \\
&& + \frac{i v g_W \lambda_T^2   P_R}{2 \sqrt{2} {m_\mathcal{T}}^4}  \biggl[
p_{b}^{\mu}\biggl(\slashed{p_h} 
+ 2  \slashed{p_W}-3m_t\biggr) 
-  p_{h}^{\mu} \biggl( \slashed{p_h}+\slashed{p_W}  -  m_t  \biggr)   \biggr]
\label{eq:frtbwh}
\end{eqnarray}

\section{$T$ Parameter in Effective Field Theory Language}
\label{sec:tparam}

The oblique parameter $\Delta{\cal {T}}$ has been calculated some time ago for the TVLQ model~\cite{Dawson:2012di}.
It is instructive to revisit this calculation using an effective field theory framework~\cite{delAguila:2016zcb} and it is an example of the importance of including the one-loop matching in SMEFT calculations.
The contributions  to $\Delta{\cal {T}}$  from fermions with masses $m_1$ and $m_2$ can be expressed in terms of the function,
\begin{eqnarray} 
\theta_+(y_1,y_2)&=&y_1+y_2-{2y_1y_2\over y_1-y_2}\log\biggl({y_1\over y_2} \biggr)-2\biggl(y_1\log(y_1)+y+2\log(y_2)\biggr)
\nonumber \\
&&-2(y_1+y_2){1\over \epsilon} \biggl({4\pi\mu^2\over M_Z^2}\biggr)^\epsilon\, ,
\end{eqnarray}
where $y_i={m_i^2 / M_Z^2}$ and $\mu$ is an arbitrary renormalization scale. 
Neglecting the $b$ quark mass and taking $m_t \gg M_Z$,  the $t-b$ contribution to the SM is found from the diagrams of Fig.~\ref{fig:tparam_sm} with SM fermion-gauge boson couplings,
\begin{eqnarray}
{\cal{T}}_{\textrm{SM}}&=& {3\over 16 \pi s_W^2 c_W^2} y_t
\end{eqnarray}
with $c_W\equiv {M_W / M_Z}$.

\begin{figure}[h]
\centering
\includegraphics[width=0.4\textwidth]{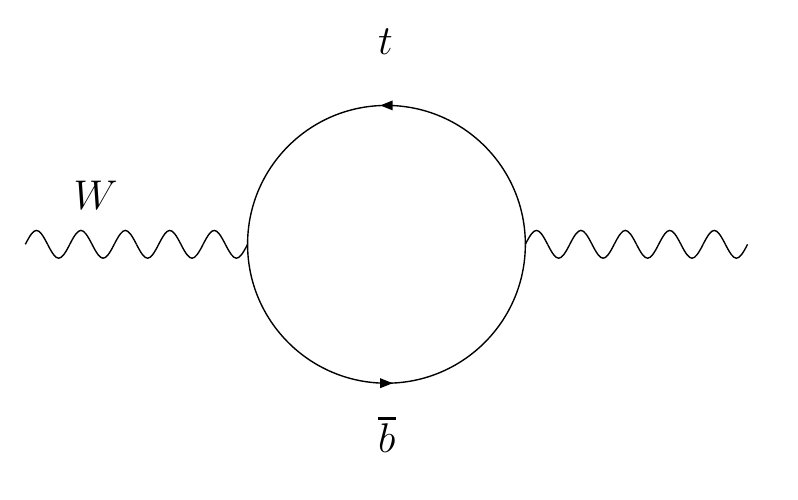}
~
\includegraphics[width=0.4\textwidth]{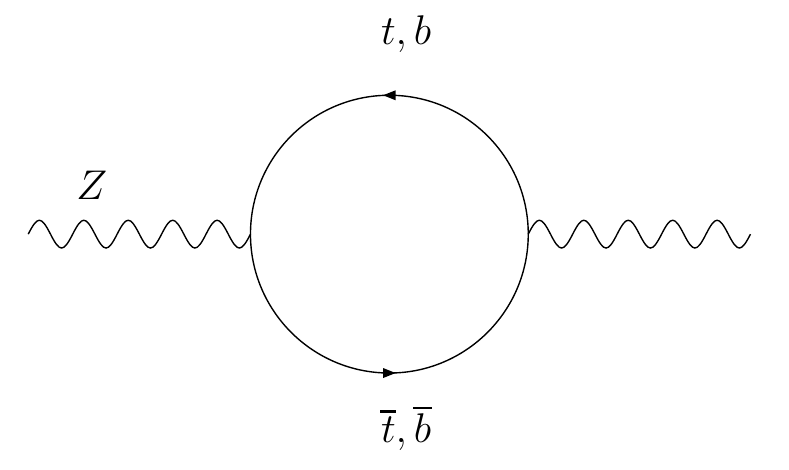}
\caption{Feynman diagrams contributing to the ${\cal{T}}$ parameter from SM fermions.}
\label{fig:tparam_sm}
\end{figure}

\begin{figure}
\centering
\includegraphics[width=0.4\textwidth]{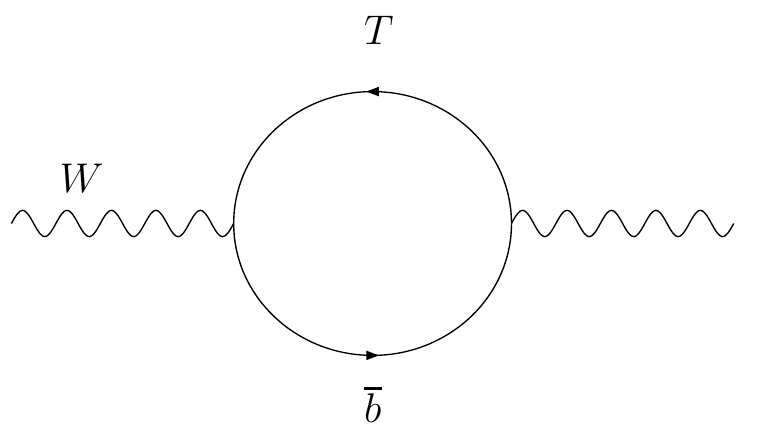}
~
\includegraphics[width=0.4\textwidth]{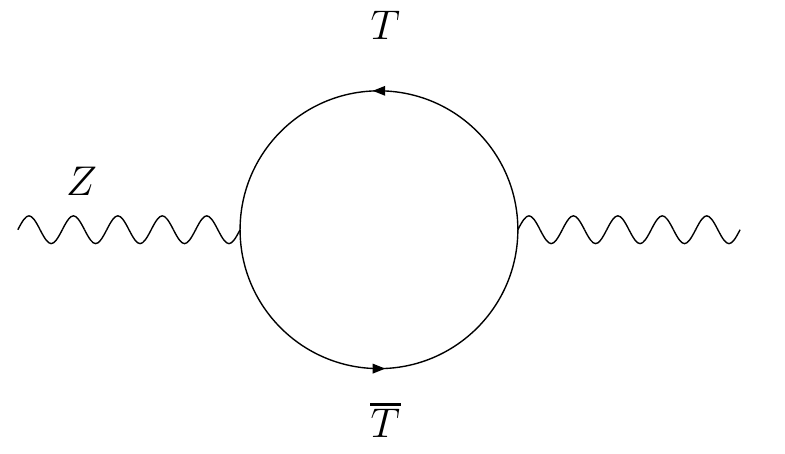}
\caption{Feynman diagrams contributing to the ${\cal{T}}$ parameter from the heavy $TVLQ$ fermion, $T$.}
\label{fig:tparam_heavy}
\end{figure}

At the UV scale (which here we take to be the physical mass of the TVLQ, $M_T$), we integrate out the contributions of the diagrams of Fig.~\ref{fig:tparam_heavy} using the couplings from Ref.~\cite{Chen:2017hak} to obtain the contribution from heavy fermions, ${\cal{T}}_H$,
\begin{eqnarray}
{\cal{T}}_H(\mu)&=& T_{\textrm{SM}}s_L^2 \biggl(\theta_+(y_T,y_b)-c_L^2 \theta_+(y_T, y_t)-{1\over 2}\theta_+(y_T,y_T)\biggr)\nonumber\\
&=& T_{\textrm{SM}} s_L^2 \biggl\{ 
s_L^2 {M_T^2\over m_t^2}-c_L^2 +{2c_L^2\over \epsilon}+2c_L^2\log\biggl(
{M_T^2\over \mu^2}\biggr)\biggr\}
\end{eqnarray}
For the UV matching, the appropriate scale is $\mu=M_T$, giving the contribution,
\begin{eqnarray}
{\cal{T}}_H(M_T)&=& T_{\textrm{SM}}s_L^2\biggl\{ 
s_L^2{M_T^2\over m_t^2}-c_L^2 +{2c_L^2\over \epsilon}\biggr\}\, .
\label{eq:theavy1}
\end{eqnarray}
Eq.~\eqref{eq:theavy1} exhibits the familiar decoupling requirement that $s_L^2M_T^2/m_t^2 \ll 1$.  

We identify,
\begin{equation}
{\cal{T}}(\mu) =-{v^2\over 2\alpha}{C^{(6)}_{HD}(\mu)\over \Lambda^2}
\end{equation}
where  $O_{HD}= | H^\dagger D_\mu H |^2$ .
The coefficient function must be renormalization group evolved to the low energy scale which we take to be $m_t$.
In the TVLQ, only the top quark Yukawa coupling contributes and we have~\cite{Jenkins:2013wua},
\begin{eqnarray}
C_{HD}(m_t)&=& C_{HD}(M_T)+{ \dot C_{HD}\over 16\pi^2}\log\biggl({m_t\over M_T}\biggr)
\end{eqnarray}
with the TVLQ result,
\begin{eqnarray}
\dot C_{HD} &=& 24\,C_{Ht}^{(1)}\, {\hat{Y}}_t^2
\end{eqnarray}
where ${\hat{Y}}_t={\sqrt{2}c_L m_t / v}$ is the top Yukawa at the UV matching scale.
Eq.~\eqref{eq:theavy1} yields,
\begin{eqnarray}
{\cal{T}}_H(m_t)&=& T_{\textrm{SM}} s_L^2\biggl\{ 
s_L^2{M_T^2\over m_t^2}-c_L^2 +{2c_L^2\over \epsilon}
-2c_L^2\log\biggl({m_t^2\over M_T^2}\biggr)\biggr\}\, .
\label{eq:theavy2}
\end{eqnarray}

Finally, at the low scale $m_t$, we need to include the diagrams of Fig.~\ref{fig:tparam_heavy} and subtract off the SM $t-b$ contribution,
\begin{eqnarray}
{\cal{T}}_L(m_t)-{\cal{T}}_{\textrm{SM}}&=& {\cal{T}}_{\textrm{SM}}s_L^2\biggl(-1-{2c_L^2\over \epsilon}\biggr) \, .
\label{eq:tlight}
\end{eqnarray}
Adding together Eqs.~\eqref{eq:theavy2} and~\eqref{eq:tlight}, we see that the effective field theory approach with with one-loop matching reproduces the correct result,
\begin{eqnarray}
\Delta {\cal{T}}&=& {\cal{T}}(m_t)_H+{\cal{T}}_L(m_t) - {\cal{T}}_{\textrm{SM}}\nonumber \\
&=& 
T_{\textrm{SM}} s_L^2 \biggl\{ 
s_L^2 {M_T^2\over m_t^2}-c_L^2 -1 +2c_L^2\log\biggl(
{M_T^2\over m_t^2}\biggr)\biggr\}\, .
\label{eq:tfinal}
\end{eqnarray}

To put Eq.~\eqref{eq:tfinal} into the language of the dimension-$6$ SMEFT, we note that
\begin{eqnarray}
s_L^2&=&{2v^2 C_{Ht}^{1,(6)}v^2\over M_T^2}+{\cal{O}}\biggl({v^4\over m_{\mathcal{T}}^4}\biggr) \nonumber \\
c_L^2&=& 1+{\cal{O}}\biggl({v^4\over m_{\mathcal{T}}^4}\biggr)\nonumber \\
s_L^4{M_T^2\over m_t^2}& =& {4[C_{Ht}^{1,(6)}]^2v^4\over M_T^2 m_t^2}
+{\cal{O}}\biggl({v^4\over m_{\mathcal{T}}^4}, {v^4\over m_{\mathcal{T}}^2m_t^2}\biggr)
\end{eqnarray}
Since we compute only single insertions of operators, Eq.~\eqref{eq:tfinal} becomes,
\begin{eqnarray}
\Delta {\cal{T}}_{\textrm{SMEFT}} &=&
2T_{\textrm{SM}}{v^2 C_{Ht}^{1,(6)}\over M_T^2}  \biggl\{ 
-2 +2\log\biggl(
{M_T^2\over m_t^2}\biggr)\biggr\} +{\cal{O}}\biggl({v^4\over m_{\mathcal{T}}^4}, {v^4\over m_{\mathcal{T}}^2m_t^2}\biggr)
\, .
\end{eqnarray}

\bibliographystyle{utphys}
\bibliography{dim8.bib}

\providecommand{\href}[2]{#2}\begingroup\raggedright\begin{thebibliography}{10}

\bibitem{Brivio:2017vri}
I.~Brivio and M.~Trott, ``{The Standard Model as an Effective Field Theory},''
  \href{http://dx.doi.org/10.1016/j.physrep.2018.11.002}{{\em Phys. Rept.}
  {\bfseries 793} (2019) 1--98},
  \href{http://arxiv.org/abs/1706.08945}{{\ttfamily arXiv:1706.08945
  [hep-ph]}}.

\bibitem{Dawson:2019clf}
S.~Dawson and P.~P. Giardino, ``{Electroweak and QCD corrections to $Z$ and $W$
  pole observables in the standard model EFT},''
  \href{http://dx.doi.org/10.1103/PhysRevD.101.013001}{{\em Phys. Rev. D}
  {\bfseries 101} no.~1, (2020) 013001},
  \href{http://arxiv.org/abs/1909.02000}{{\ttfamily arXiv:1909.02000
  [hep-ph]}}.

\bibitem{Ethier:2021bye}
J.~J. Ethier, G.~Magni, F.~Maltoni, L.~Mantani, E.~R. Nocera, J.~Rojo,
  E.~Slade, E.~Vryonidou, and C.~Zhang, ``{Combined SMEFT interpretation of
  Higgs, diboson, and top quark data from the LHC},''
  \href{http://arxiv.org/abs/2105.00006}{{\ttfamily arXiv:2105.00006
  [hep-ph]}}.

\bibitem{Almeida:2021asy}
E.~d.~S. Almeida, A.~Alves, O.~J.~P. \'Eboli, and M.~C. Gonzalez-Garcia,
  ``{Electroweak legacy of the LHC Run II},''
  \href{http://arxiv.org/abs/2108.04828}{{\ttfamily arXiv:2108.04828
  [hep-ph]}}.

\bibitem{Contino:2016jqw}
R.~Contino, A.~Falkowski, F.~Goertz, C.~Grojean, and F.~Riva, ``{On the
  Validity of the Effective Field Theory Approach to SM Precision Tests},''
  \href{http://dx.doi.org/10.1007/JHEP07(2016)144}{{\em JHEP} {\bfseries 07}
  (2016) 144}, \href{http://arxiv.org/abs/1604.06444}{{\ttfamily
  arXiv:1604.06444 [hep-ph]}}.

\bibitem{Trott:2016}
L.~Berthier and M.~Trott, ``Consistent constraints on the standard model
  effective field theory,''
  \href{http://dx.doi.org/10.1007/jhep02(2016)069}{{\em Journal of High Energy
  Physics} {\bfseries 2016} no.~2, (Feb, 2016) }.
  \url{http://dx.doi.org/10.1007/JHEP02(2016)069}.

\bibitem{Panico:2017fr}
G.~Panico, F.~Riva, and A.~Wulzer, ``{Diboson interference resurrection},''
  \href{http://dx.doi.org/10.1016/j.physletb.2017.11.068}{{\em Phys. Lett. B}
  {\bfseries 776} (2018) 473--480},
  \href{http://arxiv.org/abs/1708.07823}{{\ttfamily arXiv:1708.07823
  [hep-ph]}}.

\bibitem{Buchmuller:1985jz}
W.~Buchmuller and D.~Wyler, ``{Effective Lagrangian Analysis of New
  Interactions and Flavor Conservation},''
  \href{http://dx.doi.org/10.1016/0550-3213(86)90262-2}{{\em Nucl. Phys. B}
  {\bfseries 268} (1986) 621--653}.

\bibitem{ZZ:2020}
J.~Ellis, S.-F. Ge, H.-J. He, and R.-Q. Xiao, ``Probing the scale of new
  physics in the zz$\gamma$ coupling at $e^+e^-$ colliders,''
  \href{http://dx.doi.org/10.1088/1674-1137/44/6/063106}{{\em Chinese Physics
  C} {\bfseries 44} no.~6, (Jun, 2020) 063106}.
  \url{http://dx.doi.org/10.1088/1674-1137/44/6/063106}.

\bibitem{Hays:2018zze}
C.~Hays, A.~Martin, V.~Sanz, and J.~Setford, ``{On the impact of
  dimension-eight SMEFT operators on Higgs measurements},''
  \href{http://dx.doi.org/10.1007/JHEP02(2019)123}{{\em JHEP} {\bfseries 02}
  (2019) 123}, \href{http://arxiv.org/abs/1808.00442}{{\ttfamily
  arXiv:1808.00442 [hep-ph]}}.

\bibitem{Murphy:2020ab}
C.~W. Murphy, ``{Dimension-8 operators in the Standard Model Effective Field
  Theory},'' \href{http://dx.doi.org/10.1007/jhep10(2020)174}{{\em Journal of
  High Energy Physics} {\bfseries 2020} no.~10, (Oct, 2020) }.
  \url{http://dx.doi.org/10.1007/JHEP10(2020)174}.

\bibitem{Murphy:2020cly}
C.~W. Murphy, ``{Low-Energy Effective Field Theory below the Electroweak Scale:
  Dimension-8 Operators},''
  \href{http://dx.doi.org/10.1007/JHEP04(2021)101}{{\em JHEP} {\bfseries 04}
  (2021) 101}, \href{http://arxiv.org/abs/2012.13291}{{\ttfamily
  arXiv:2012.13291 [hep-ph]}}.

\bibitem{Li:2020tsi}
H.-L. Li, Z.~Ren, M.-L. Xiao, J.-H. Yu, and Y.-H. Zheng, ``{Low energy
  effective field theory operator basis at d \ensuremath{\leq} 9},''
  \href{http://dx.doi.org/10.1007/JHEP06(2021)138}{{\em JHEP} {\bfseries 06}
  (2021) 138}, \href{http://arxiv.org/abs/2012.09188}{{\ttfamily
  arXiv:2012.09188 [hep-ph]}}.

\bibitem{Lin:2021}
H.-L. Li, Z.~Ren, J.~Shu, M.-L. Xiao, J.-H. Yu, and Y.-H. Zheng, ``Complete set
  of dimension-eight operators in the standard model effective field theory,''
  \href{http://dx.doi.org/10.1103/physrevd.104.015026}{{\em Physical Review D}
  {\bfseries 104} no.~1, (Jul, 2021) }.
  \url{http://dx.doi.org/10.1103/PhysRevD.104.015026}.

\bibitem{Dawson:2015gka}
S.~Dawson, I.~M. Lewis, and M.~Zeng, ``{Usefulness of effective field theory
  for boosted Higgs production},''
  \href{http://dx.doi.org/10.1103/PhysRevD.91.074012}{{\em Phys. Rev. D}
  {\bfseries 91} (2015) 074012},
  \href{http://arxiv.org/abs/1501.04103}{{\ttfamily arXiv:1501.04103
  [hep-ph]}}.

\bibitem{Dawson:2014ora}
S.~Dawson, I.~M. Lewis, and M.~Zeng, ``{Effective field theory for Higgs boson
  plus jet production},''
  \href{http://dx.doi.org/10.1103/PhysRevD.90.093007}{{\em Phys. Rev. D}
  {\bfseries 90} no.~9, (2014) 093007},
  \href{http://arxiv.org/abs/1409.6299}{{\ttfamily arXiv:1409.6299 [hep-ph]}}.

\bibitem{Harlander:2013oja}
R.~V. Harlander and T.~Neumann, ``{Probing the nature of the Higgs-gluon
  coupling},'' \href{http://dx.doi.org/10.1103/PhysRevD.88.074015}{{\em Phys.
  Rev. D} {\bfseries 88} (2013) 074015},
  \href{http://arxiv.org/abs/1308.2225}{{\ttfamily arXiv:1308.2225 [hep-ph]}}.

\bibitem{Grazzini:2016paz}
M.~Grazzini, A.~Ilnicka, M.~Spira, and M.~Wiesemann, ``{Modeling BSM effects on
  the Higgs transverse-momentum spectrum in an EFT approach},''
  \href{http://dx.doi.org/10.1007/JHEP03(2017)115}{{\em JHEP} {\bfseries 03}
  (2017) 115}, \href{http://arxiv.org/abs/1612.00283}{{\ttfamily
  arXiv:1612.00283 [hep-ph]}}.

\bibitem{Battaglia:2021nys}
M.~Battaglia, M.~Grazzini, M.~Spira, and M.~Wiesemann, ``{Sensitivity to BSM
  effects in the Higgs $p_T$ spectrum within SMEFT},''
  \href{http://arxiv.org/abs/2109.02987}{{\ttfamily arXiv:2109.02987
  [hep-ph]}}.

\bibitem{Corbett:2021eux}
T.~Corbett, A.~Helset, A.~Martin, and M.~Trott, ``{EWPD in the SMEFT to
  dimension eight},'' \href{http://dx.doi.org/10.1007/JHEP06(2021)076}{{\em
  JHEP} {\bfseries 06} (2021) 076},
  \href{http://arxiv.org/abs/2102.02819}{{\ttfamily arXiv:2102.02819
  [hep-ph]}}.

\bibitem{Dawson:2020oco}
S.~Dawson, S.~Homiller, and S.~D. Lane, ``{Putting standard model EFT fits to
  work},'' \href{http://dx.doi.org/10.1103/PhysRevD.102.055012}{{\em Phys. Rev.
  D} {\bfseries 102} no.~5, (2020) 055012},
  \href{http://arxiv.org/abs/2007.01296}{{\ttfamily arXiv:2007.01296
  [hep-ph]}}.

\bibitem{Brivio:2021alv}
I.~Brivio, S.~Bruggisser, E.~Geoffray, W.~Kilian, M.~Kr\"amer, M.~Luchmann,
  T.~Plehn, and B.~Summ, ``{From Models to SMEFT and Back?},''
  \href{http://arxiv.org/abs/2108.01094}{{\ttfamily arXiv:2108.01094
  [hep-ph]}}.

\bibitem{Arkani-Hamed:2002ikv}
N.~Arkani-Hamed, A.~G. Cohen, E.~Katz, and A.~E. Nelson, ``{The Littlest
  Higgs},'' \href{http://dx.doi.org/10.1088/1126-6708/2002/07/034}{{\em JHEP}
  {\bfseries 07} (2002) 034},
  \href{http://arxiv.org/abs/hep-ph/0206021}{{\ttfamily arXiv:hep-ph/0206021}}.

\bibitem{Perelstein:2003wd}
M.~Perelstein, M.~E. Peskin, and A.~Pierce, ``{Top quarks and electroweak
  symmetry breaking in little Higgs models},''
  \href{http://dx.doi.org/10.1103/PhysRevD.69.075002}{{\em Phys. Rev. D}
  {\bfseries 69} (2004) 075002},
  \href{http://arxiv.org/abs/hep-ph/0310039}{{\ttfamily arXiv:hep-ph/0310039}}.

\bibitem{Csaki:2002qg}
C.~Csaki, J.~Hubisz, G.~D. Kribs, P.~Meade, and J.~Terning, ``{Big corrections
  from a little Higgs},''
  \href{http://dx.doi.org/10.1103/PhysRevD.67.115002}{{\em Phys. Rev. D}
  {\bfseries 67} (2003) 115002},
  \href{http://arxiv.org/abs/hep-ph/0211124}{{\ttfamily arXiv:hep-ph/0211124}}.

\bibitem{Panico:2015jxa}
G.~Panico and A.~Wulzer,
  \href{http://dx.doi.org/10.1007/978-3-319-22617-0}{{\em {The Composite
  Nambu-Goldstone Higgs}}}, vol.~913.
\newblock Springer, 2016.
\newblock \href{http://arxiv.org/abs/1506.01961}{{\ttfamily arXiv:1506.01961
  [hep-ph]}}.

\bibitem{Matsedonskyi:2015dns}
O.~Matsedonskyi, G.~Panico, and A.~Wulzer, ``{Top Partners Searches and
  Composite Higgs Models},''
  \href{http://dx.doi.org/10.1007/JHEP04(2016)003}{{\em JHEP} {\bfseries 04}
  (2016) 003}, \href{http://arxiv.org/abs/1512.04356}{{\ttfamily
  arXiv:1512.04356 [hep-ph]}}.

\bibitem{Dobrescu_1998}
B.~A. Dobrescu and C.~T. Hill, ``Electroweak symmetry breaking via a top
  condensation seesaw mechanism,''
  \href{http://dx.doi.org/10.1103/physrevlett.81.2634}{{\em Physical Review
  Letters} {\bfseries 81} no.~13, (Sep, 1998) 2634?2637}.
  \url{http://dx.doi.org/10.1103/PhysRevLett.81.2634}.

\bibitem{He_2002}
H.-J. He, C.~T. Hill, and T.~M.~P. Tait, ``Top quark seesaw model, vacuum
  structure, and electroweak precision constraints,''
  \href{http://dx.doi.org/10.1103/physrevd.65.055006}{{\em Physical Review D}
  {\bfseries 65} no.~5, (Feb, 2002) }.
  \url{http://dx.doi.org/10.1103/PhysRevD.65.055006}.

\bibitem{Gaillard:1985uh}
M.~K. Gaillard, ``{The Effective One Loop Lagrangian With Derivative
  Couplings},'' \href{http://dx.doi.org/10.1016/0550-3213(86)90264-6}{{\em
  Nucl. Phys. B} {\bfseries 268} (1986) 669--692}.

\bibitem{Henning:2014wua}
B.~Henning, X.~Lu, and H.~Murayama, ``{How to use the Standard Model effective
  field theory},'' \href{http://dx.doi.org/10.1007/JHEP01(2016)023}{{\em JHEP}
  {\bfseries 01} (2016) 023}, \href{http://arxiv.org/abs/1412.1837}{{\ttfamily
  arXiv:1412.1837 [hep-ph]}}.

\bibitem{Beenakker:2001rj}
W.~Beenakker, S.~Dittmaier, M.~Kramer, B.~Plumper, M.~Spira, and P.~M. Zerwas,
  ``{Higgs radiation off top quarks at the Tevatron and the LHC},''
  \href{http://dx.doi.org/10.1103/PhysRevLett.87.201805}{{\em Phys. Rev. Lett.}
  {\bfseries 87} (2001) 201805},
  \href{http://arxiv.org/abs/hep-ph/0107081}{{\ttfamily arXiv:hep-ph/0107081}}.

\bibitem{Beenakker:2002nc}
W.~Beenakker, S.~Dittmaier, M.~Kramer, B.~Plumper, M.~Spira, and P.~M. Zerwas,
  ``{NLO QCD corrections to t anti-t H production in hadron collisions},''
  \href{http://dx.doi.org/10.1016/S0550-3213(03)00044-0}{{\em Nucl. Phys. B}
  {\bfseries 653} (2003) 151--203},
  \href{http://arxiv.org/abs/hep-ph/0211352}{{\ttfamily arXiv:hep-ph/0211352}}.

\bibitem{Dawson:2002tg}
S.~Dawson, L.~H. Orr, L.~Reina, and D.~Wackeroth, ``{Associated top quark Higgs
  boson production at the LHC},''
  \href{http://dx.doi.org/10.1103/PhysRevD.67.071503}{{\em Phys. Rev. D}
  {\bfseries 67} (2003) 071503},
  \href{http://arxiv.org/abs/hep-ph/0211438}{{\ttfamily arXiv:hep-ph/0211438}}.

\bibitem{Dawson:2003zu}
S.~Dawson, C.~Jackson, L.~H. Orr, L.~Reina, and D.~Wackeroth, ``{Associated
  Higgs production with top quarks at the large hadron collider: NLO QCD
  corrections},'' \href{http://dx.doi.org/10.1103/PhysRevD.68.034022}{{\em
  Phys. Rev. D} {\bfseries 68} (2003) 034022},
  \href{http://arxiv.org/abs/hep-ph/0305087}{{\ttfamily arXiv:hep-ph/0305087}}.

\bibitem{Cacciapaglia:2010vn}
G.~Cacciapaglia, A.~Deandrea, D.~Harada, and Y.~Okada, ``{Bounds and Decays of
  New Heavy Vector-like Top Partners},''
  \href{http://dx.doi.org/10.1007/JHEP11(2010)159}{{\em JHEP} {\bfseries 11}
  (2010) 159}, \href{http://arxiv.org/abs/1007.2933}{{\ttfamily arXiv:1007.2933
  [hep-ph]}}.

\bibitem{Chen:2017hak}
C.-Y. Chen, S.~Dawson, and E.~Furlan, ``{Vectorlike fermions and Higgs
  effective field theory revisited},''
  \href{http://dx.doi.org/10.1103/PhysRevD.96.015006}{{\em Phys. Rev. D}
  {\bfseries 96} no.~1, (2017) 015006},
  \href{http://arxiv.org/abs/1703.06134}{{\ttfamily arXiv:1703.06134
  [hep-ph]}}.

\bibitem{Cacciapaglia:2018qep}
G.~Cacciapaglia, A.~Carvalho, A.~Deandrea, T.~Flacke, B.~Fuks, D.~Majumder,
  L.~Panizzi, and H.-S. Shao, ``{Next-to-leading-order predictions for single
  vector-like quark production at the LHC},''
  \href{http://dx.doi.org/10.1016/j.physletb.2019.04.056}{{\em Phys. Lett. B}
  {\bfseries 793} (2019) 206--211},
  \href{http://arxiv.org/abs/1811.05055}{{\ttfamily arXiv:1811.05055
  [hep-ph]}}.

\bibitem{Buchkremer:2013bha}
M.~Buchkremer, G.~Cacciapaglia, A.~Deandrea, and L.~Panizzi, ``{Model
  Independent Framework for Searches of Top Partners},''
  \href{http://dx.doi.org/10.1016/j.nuclphysb.2013.08.010}{{\em Nucl. Phys. B}
  {\bfseries 876} (2013) 376--417},
  \href{http://arxiv.org/abs/1305.4172}{{\ttfamily arXiv:1305.4172 [hep-ph]}}.

\bibitem{Cacciapaglia:2011fx}
G.~Cacciapaglia, A.~Deandrea, L.~Panizzi, N.~Gaur, D.~Harada, and Y.~Okada,
  ``{Heavy Vector-like Top Partners at the LHC and flavour constraints},''
  \href{http://dx.doi.org/10.1007/JHEP03(2012)070}{{\em JHEP} {\bfseries 03}
  (2012) 070}, \href{http://arxiv.org/abs/1108.6329}{{\ttfamily arXiv:1108.6329
  [hep-ph]}}.

\bibitem{Matsedonskyi:2014mna}
O.~Matsedonskyi, G.~Panico, and A.~Wulzer, ``{On the Interpretation of Top
  Partners Searches},'' \href{http://dx.doi.org/10.1007/JHEP12(2014)097}{{\em
  JHEP} {\bfseries 12} (2014) 097},
  \href{http://arxiv.org/abs/1409.0100}{{\ttfamily arXiv:1409.0100 [hep-ph]}}.

\bibitem{delAguila:1998tp}
F.~del Aguila, J.~A. Aguilar-Saavedra, and R.~Miquel, ``{Constraints on top
  couplings in models with exotic quarks},''
  \href{http://dx.doi.org/10.1103/PhysRevLett.82.1628}{{\em Phys. Rev. Lett.}
  {\bfseries 82} (1999) 1628--1631},
  \href{http://arxiv.org/abs/hep-ph/9808400}{{\ttfamily arXiv:hep-ph/9808400}}.

\bibitem{Aguilar-Saavedra:2002phh}
J.~A. Aguilar-Saavedra, ``{Effects of mixing with quark singlets},''
  \href{http://dx.doi.org/10.1103/PhysRevD.69.099901}{{\em Phys. Rev. D}
  {\bfseries 67} (2003) 035003},
  \href{http://arxiv.org/abs/hep-ph/0210112}{{\ttfamily arXiv:hep-ph/0210112}}.
  [Erratum: Phys.Rev.D 69, 099901 (2004)].

\bibitem{Aguilar-Saavedra:2013qpa}
J.~A. Aguilar-Saavedra, R.~Benbrik, S.~Heinemeyer, and M.~P\'erez-Victoria,
  ``{Handbook of vectorlike quarks: Mixing and single production},''
  \href{http://dx.doi.org/10.1103/PhysRevD.88.094010}{{\em Phys. Rev. D}
  {\bfseries 88} no.~9, (2013) 094010},
  \href{http://arxiv.org/abs/1306.0572}{{\ttfamily arXiv:1306.0572 [hep-ph]}}.

\bibitem{Ellis:2014dza}
S.~A.~R. Ellis, R.~M. Godbole, S.~Gopalakrishna, and J.~D. Wells, ``{Survey of
  vector-like fermion extensions of the Standard Model and their
  phenomenological implications},''
  \href{http://dx.doi.org/10.1007/JHEP09(2014)130}{{\em JHEP} {\bfseries 09}
  (2014) 130}, \href{http://arxiv.org/abs/1404.4398}{{\ttfamily arXiv:1404.4398
  [hep-ph]}}.

\bibitem{Aguila_2000}
F.~d. Aguila, J.~Santiago, and M.~Perez-Victoria, ``Observable contributions of
  new exotic quarks to quark mixing,''
  \href{http://dx.doi.org/10.1088/1126-6708/2000/09/011}{{\em Journal of High
  Energy Physics} {\bfseries 2000} no.~09, (Sep, 2000) 011?011}.
  \url{http://dx.doi.org/10.1088/1126-6708/2000/09/011}.

\bibitem{Chen:2014xwa}
C.-Y. Chen, S.~Dawson, and I.~M. Lewis, ``{Top Partners and Higgs Boson
  Production},'' \href{http://dx.doi.org/10.1103/PhysRevD.90.035016}{{\em Phys.
  Rev. D} {\bfseries 90} no.~3, (2014) 035016},
  \href{http://arxiv.org/abs/1406.3349}{{\ttfamily arXiv:1406.3349 [hep-ph]}}.

\bibitem{Buchkremer_2013}
M.~Buchkremer, G.~Cacciapaglia, A.~Deandrea, and L.~Panizzi,
  ``Model-independent framework for searches of top partners,''
  \href{http://dx.doi.org/10.1016/j.nuclphysb.2013.08.010}{{\em Nuclear Physics
  B} {\bfseries 876} no.~2, (Nov, 2013) 376?417}.
  \url{http://dx.doi.org/10.1016/j.nuclphysb.2013.08.010}.

\bibitem{Chanowitz:1978mv}
M.~S. Chanowitz, M.~A. Furman, and I.~Hinchliffe, ``{Weak Interactions of
  Ultraheavy Fermions. 2.},''
  \href{http://dx.doi.org/10.1016/0550-3213(79)90606-0}{{\em Nucl. Phys. B}
  {\bfseries 153} (1979) 402--430}.

\bibitem{ATLAS:2018ziw}
{\bfseries ATLAS} Collaboration, M.~Aaboud {\em et~al.}, ``{Combination of the
  searches for pair-produced vector-like partners of the third-generation
  quarks at $\sqrt{s} =$ 13 TeV with the ATLAS detector},''
  \href{http://dx.doi.org/10.1103/PhysRevLett.121.211801}{{\em Phys. Rev.
  Lett.} {\bfseries 121} no.~21, (2018) 211801},
  \href{http://arxiv.org/abs/1808.02343}{{\ttfamily arXiv:1808.02343
  [hep-ex]}}.

\bibitem{cms_2018}
A.~M. Sirunyan, A.~Tumasyan, W.~Adam, F.~Ambrogi, E.~Asilar, T.~Bergauer,
  J.~Brandstetter, E.~Brondolin, M.~Dragicevic, and et~al., ``Search for
  vector-like t and b quark pairs in final states with leptons at $ \sqrt{s}=13
  $ tev,'' \href{http://dx.doi.org/10.1007/jhep08(2018)177}{{\em Journal of
  High Energy Physics} {\bfseries 2018} no.~8, (Aug, 2018) }.
  \url{http://dx.doi.org/10.1007/JHEP08(2018)177}.

\bibitem{Liu_2019}
Y.-B. Liu and S.~Moretti, ``Search for single production of a top quark partner
  via the t$\rightarrow th$ and $h\rightarrow$ ww* channels at the lhc,''
  \href{http://dx.doi.org/10.1103/physrevd.100.015025}{{\em Physical Review D}
  {\bfseries 100} no.~1, (Jul, 2019) }.
  \url{http://dx.doi.org/10.1103/PhysRevD.100.015025}.

\bibitem{Yang:2021btv}
B.~Yang, M.~Wang, H.~Bi, and L.~Shang, ``{Single production of vectorlike $T$
  quark decaying into $Wb$ at the LHC and the future $pp$ colliders},''
  \href{http://dx.doi.org/10.1103/PhysRevD.103.036006}{{\em Phys. Rev. D}
  {\bfseries 103} no.~3, (2021) 036006}.

\bibitem{ATLAS_singt}
{\bfseries ATLAS} Collaboration, M.~Aaboud {\em et~al.}, ``{Search for single
  production of vector-like T quarks decaying to Ht or Zt in pp collisions
  $\sqrt{s} =$ 13 TeV with the ATLAS detector},''.
  \url{http://cdsweb.cern.ch/record/2779174/files/ATLAS-CONF-2021-040.pdf}.

\bibitem{Buras_2011}
A.~J. Buras, C.~Grojean, S.~Pokorski, and R.~Ziegler, ``Fcnc effects in a
  minimal theory of fermion masses,''
  \href{http://dx.doi.org/10.1007/jhep08(2011)028}{{\em Journal of High Energy
  Physics} {\bfseries 2011} no.~8, (Aug, 2011) }.
  \url{http://dx.doi.org/10.1007/JHEP08(2011)028}.

\bibitem{Criado:2018sdb}
J.~C. Criado and M.~P\'erez-Victoria, ``{Field redefinitions in effective
  theories at higher orders},''
  \href{http://dx.doi.org/10.1007/JHEP03(2019)038}{{\em JHEP} {\bfseries 03}
  (2019) 038}, \href{http://arxiv.org/abs/1811.09413}{{\ttfamily
  arXiv:1811.09413 [hep-ph]}}.

\bibitem{Brivio:2017bnu}
I.~Brivio and M.~Trott, ``{Scheming in the SMEFT... and a reparameterization
  invariance!},'' \href{http://dx.doi.org/10.1007/JHEP07(2017)148}{{\em JHEP}
  {\bfseries 07} (2017) 148}, \href{http://arxiv.org/abs/1701.06424}{{\ttfamily
  arXiv:1701.06424 [hep-ph]}}. [Addendum: JHEP 05, 136 (2018)].

\bibitem{Degrande_2012}
C.~Degrande, C.~Duhr, B.~Fuks, D.~Grellscheid, O.~Mattelaer, and T.~Reiter,
  ``Ufo- the universal feynrules output,''
  \href{http://dx.doi.org/10.1016/j.cpc.2012.01.022}{{\em Computer Physics
  Communications} {\bfseries 183} no.~6, (Jun, 2012) 1201--1214}.
  \url{http://dx.doi.org/10.1016/j.cpc.2012.01.022}.

\bibitem{Alloul_2014}
A.~Alloul, N.~D. Christensen, C.~Degrande, C.~Duhr, and B.~Fuks,
  ``Feynrules2.0, a complete toolbox for tree level phenomenology,''
  \href{http://dx.doi.org/10.1016/j.cpc.2014.04.012}{{\em Computer Physics
  Communications} {\bfseries 185} no.~8, (Aug, 2014) 2250--2300}.
  \url{http://dx.doi.org/10.1016/j.cpc.2014.04.012}.

\bibitem{Grzadkowski_2010}
B.~Grzadkowski, M.~Iskrzy?ski, M.~Misiak, and J.~Rosiek, ``Dimension-six terms
  in the standard model lagrangian,''
  \href{http://dx.doi.org/10.1007/jhep10(2010)085}{{\em Journal of High Energy
  Physics} {\bfseries 2010} no.~10, (Oct, 2010) }.
  \url{http://dx.doi.org/10.1007/JHEP10(2010)085}.

\bibitem{Brehmer:2015rna}
J.~Brehmer, A.~Freitas, D.~Lopez-Val, and T.~Plehn, ``{Pushing Higgs Effective
  Theory to its Limits},''
  \href{http://dx.doi.org/10.1103/PhysRevD.93.075014}{{\em Phys. Rev. D}
  {\bfseries 93} no.~7, (2016) 075014},
  \href{http://arxiv.org/abs/1510.03443}{{\ttfamily arXiv:1510.03443
  [hep-ph]}}.

\bibitem{Egana-Ugrinovic:2015vgy}
D.~Egana-Ugrinovic and S.~Thomas, ``{Effective Theory of Higgs Sector Vacuum
  States},'' \href{http://arxiv.org/abs/1512.00144}{{\ttfamily arXiv:1512.00144
  [hep-ph]}}.

\bibitem{B_lusca_Ma_to_2017}
H.~Belusca-Maito, A.~Falkowski, D.~Fontes, J.~C. Romao, and J.~P. Silva,
  ``Higgs eft for 2hdm and beyond,''
  \href{http://dx.doi.org/10.1140/epjc/s10052-017-4745-5}{{\em The European
  Physical Journal C} {\bfseries 77} no.~3, (Mar, 2017) }.
  \url{http://dx.doi.org/10.1140/epjc/s10052-017-4745-5}.

\bibitem{Dawson:2012di}
S.~Dawson and E.~Furlan, ``{A Higgs Conundrum with Vector Fermions},''
  \href{http://dx.doi.org/10.1103/PhysRevD.86.015021}{{\em Phys. Rev. D}
  {\bfseries 86} (2012) 015021},
  \href{http://arxiv.org/abs/1205.4733}{{\ttfamily arXiv:1205.4733 [hep-ph]}}.

\bibitem{Christensen_2009}
N.~D. Christensen and C.~Duhr, ``Feynrules: Feynman rules made easy,''
  \href{http://dx.doi.org/10.1016/j.cpc.2009.02.018}{{\em Computer Physics
  Communications} {\bfseries 180} no.~9, (Sep, 2009) 1614--1641}.
  \url{http://dx.doi.org/10.1016/j.cpc.2009.02.018}.

\bibitem{Corbett:2021iob}
T.~Corbett and T.~Rasmussen, ``{Higgs decays to two leptons and a photon beyond
  leading order in the SMEFT},''
  \href{http://arxiv.org/abs/2110.03694}{{\ttfamily arXiv:2110.03694
  [hep-ph]}}.

\bibitem{delAguila:2016zcb}
F.~del Aguila, Z.~Kunszt, and J.~Santiago, ``{One-loop effective lagrangians
  after matching},''
  \href{http://dx.doi.org/10.1140/epjc/s10052-016-4081-1}{{\em Eur. Phys. J. C}
  {\bfseries 76} no.~5, (2016) 244},
  \href{http://arxiv.org/abs/1602.00126}{{\ttfamily arXiv:1602.00126
  [hep-ph]}}.

\bibitem{Jenkins:2013wua}
E.~E. Jenkins, A.~V. Manohar, and M.~Trott, ``{Renormalization Group Evolution
  of the Standard Model Dimension Six Operators II: Yukawa Dependence},''
  \href{http://dx.doi.org/10.1007/JHEP01(2014)035}{{\em JHEP} {\bfseries 01}
  (2014) 035}, \href{http://arxiv.org/abs/1310.4838}{{\ttfamily arXiv:1310.4838
  [hep-ph]}}.

\end{thebibliography}\endgroup

\end{document}